\documentclass[runningheads]{svmult}

\usepackage{makeidx}   
\usepackage{graphicx}  
\usepackage{subeqnar}  
\usepackage{multicol}  
\usepackage{physprbb}  
\makeindex             

\begin{document}
\title*{The tensor virial method and its applications to 
self-gravitating superfluids}
\toctitle{The tensor virial method and its applications to 
self-gravitating superfluids}
%
%
\titlerunning{The tensor virial method}
%
\author{A. Sedrakian\inst{1,2}
\and I. Wasserman\inst{3}
\authorrunning{Sedrakian and Wasserman}
%
%
\institute{ Kernfysisch Versneller Instituut,\\
            Groningen AA-9747, \\
            The Netherlands\\
\and Groupe de Physique Theorique,\\
     Institut de Physique Nucleaire,\\
     F-91406 Orsay Cedex\\
\and Center for Radiophysics and Space Research,\\
     Cornell University,\\
     Ithaca, NY 14853     
     }
     }

\maketitle              

\begin{abstract}
This review starts with a discussion of the hierarchy of scales, 
relevant to the description of superfluids in neutron stars, 
which motivates a subsequent elementary exposition of the Newtonian 
superfluid hydrodynamics. Starting from the Euler equations 
for a superfluid and a normal fluid we apply the tensor virial method 
to obtain the virial equations of the first, second, and third order
and to compute their Eulerian perturbations. Special emphasis is put 
on the computation of perturbations of the new terms 
due to mutual gravitational attraction and mutual 
friction between the two fluids. The oscillation modes of superfluid 
Maclaurin spheroids are derived from the first and second order perturbed 
virial equations. We discuss two generic classes 
of oscillation modes which correspond to the {\it co-moving} 
and {\it relative oscillations} of two fluids. These modes decouple if 
the normal fluid is inviscid. We also discuss the mixing of 
these modes (when the  normal fluid is viscous) 
and its effect on the dynamical and secular instabilities of 
the co-moving modes and their damping.
\end{abstract}

\section{Introduction}

Radio and x-ray observations of neutron stars
provide strong evidence for the 
superfluidity of neutron star interiors.
Perhaps, the most striking manifestations of their
superfluidity are the long (on time-scales from several 
hours to hundreds of days) relaxations that follow the glitches in 
the spin and spin-down rates of some pulsars.
Although the majority of pulsars are very precise clocks, 
timing observations reveal persistent random fluctuations in 
times of arrival of radio signals. Some pulsars show
long-term periodicities in their spin-characteristics 
and periodic changes of their pulse shape. However, the 
relation between the superfluidity of neutron star interiors 
and these anomalies of pulsar timing is not firmly established 
yet.

Further evidence for the superfluidity of neutron star interiors
came in the 1980s with the advent of the orbiting x-ray satellites.
The measurements of the thermal radiation from a dozen or so hot 
neutron stars provided indirect information on the temperature
of superfluid phases in neutron stars.
Theoretical thermal histories of superfluid neutron stars
are consistent with the x-ray data (within the limits of 
our knowledge of the input physics).
Non-superfluid stars, as a rule, cool too fast to 
below the threshold of detection.

Apart from the radiation in the electro-magnetic spectrum, 
neutron stars are expected to be primary sources of 
gravitational wave radiation, which are expected to be detectable by 
future laser interferometer detectors. It is hoped that 
one can probe neutron star interiors and their superfluidity 
using gravity waves, as their eigen-frequencies and  
damping may depend on the dissipation in the superfluid, 
at temperatures below the superfluid phase transition.

This review concentrates on the oscillations of superfluid 
self-gravitating ellipsoids within the tensor virial method.
The method was originally introduced by Chandrasekhar and Fermi 
in the context of magneto-hydrodynamics \cite{CHANDRA_FERMI}. 
It was extensively developed   in the 1960s
by Chandrasekhar for the study of the ellipsoidal figures 
of equilibrium and their oscillations. A comprehensive account 
of this work is contained in Chandrasekhar's monograph 
{\it Ellipsoidal Figures of Equilibrium} (hereafter EFE)\cite{CHANDRA}.

The ellipsoidal approximation provides an idealized picture 
of oscillations of neutron stars. One can think of several arguments 
in favor of adopting such an approach: first, the combination of the 
Newtonian gravity and  two-fluid hydrodynamics of superfluid  defines
an exactly solvable model if we assume that the fluids are
incompressible and inviscid; second, past experience with single-fluid 
self-gravitating ellipsoids shows that most of the qualitative features 
found for these ellipsoids have their analogs in more ``realistic''
systems; third, the method is transparent and in many cases analytical
results can be obtained which shed light on the underlying physics. 

The tensor virial method is not the only  tool 
for investigating the properties of ellipsoidal figures.
Alternative formulations exist in the literature
and we refer to Refs. \cite{CARTER,IPSER,LRS} for further details;
for a pedagogical introduction see the textbook \cite{ST}. 
Note, also,
that various formulations of the theory of ellipsoids,
to a large extent, are equivalent. 

Superfluid oscillations were studied using various methods and 
approximations in the past decade. Epstein pointed out that the 
superfluidity of neutron star interiors has potentially 
important effects on the propagation of seismically excited
acoustic waves. It allows for additional types of waves to propagate
by virtue of doubling of degrees of freedom in a superfluid; 
superfluid phases create acoustic discontinuities 
in which wave velocities or polarizations change abruptly on the 
bounding interfaces \cite{EPSTEIN}.
The effects of superfluidity on global
hydrodynamic oscillations were investigated
by Lindblom and Mendell \cite{LINDBLOM1} in a model where the superfluid
and the normal fluid are coupled via  gravitational attraction 
and the {\it entrainment effect}\footnote{The latter effect 
arises in the layers where  neutrons and proton condensates coexist, hence 
the flow of one condensate is accompanied by the motion 
of the other. See B.~Carter's article in this volume for further
details.}. 
Their solutions reveal that the lowest frequency pulsations are 
almost indistinguishable from those derived from the ordinary-fluid 
hydrodynamics; however, their analytical solutions also
reveal the existence of a 
spectrum of modes which are absent in a single fluid star. 
Nonradial oscillations of non-rotating superfluid
neutron stars were computed by Lee, whose numerical solutions for 
the radial and non-radial pulsations of the two-fluid stars identified 
distinct superfluid modes \cite{UMIN}.
The effects of shear viscosity of the electron fluid and mutual 
friction on the r-mode oscillations were studied by Lindblom and 
Mendell \cite{LINDBLOM2} by constructing an energy functional
and computing the time-scales associated with the dissipative 
terms. The oscillation modes of superfluid analogs of the 
classical Maclaurin, Jacobi and Roche ellipsoids were derived recently 
by Sedrakian and Wasserman \cite{SW2000} within the 
tensor virial method. The oscillation modes of superfluid 
ellipsoids separate into two classes corresponding to 
relative and co-moving (or center-of-mass,  hereafter referred as CM)
motions of two fluids. The 
CM oscillations are identical to the oscillations 
of single-fluid ellipsoids and 
are undamped if one ignores the viscosity of
the normal fluid. The mutual friction contributes only to
the damping of the relative oscillation modes. One important 
feature of the latter modes is that they do not emit gravitational
radiation as there is no mass transport associated with them.
Our  discussion of the tensor virial method is based 
on Ref. \cite{SW2000}. 

This review is organized as follows. In the remainder of the Introduction
we motivate the averaged superfluid hydrodynamics and identify the relevant 
scales in the problem. In Sect. 2 we give  a tutorial introduction 
to the Newtonian superfluid hydrodynamics, 
which is mainly built on the work of Bekarevich and 
Khalatnikov \cite{BK}. Sect. 3 introduces
the tensor virial method 
and illustrates its applications to the superfluid ellipsoidal figures 
by computing the perturbations of the new terms in virial equations due 
to their two-fluid nature. We discuss 
in Sect. 4  the oscillations of superfluid Maclaurin spheroids 
including the effects of mutual friction 
and viscosity of the normal fluid. Sect. 5 is a brief summary.
We refer the reader to the accompanying article by B.~Carter 
for a review of relativistic models and an overview of 
the state of the art of the theory of superfluidity 
in neutron stars.

\subsection{Characteristic length scales}

The physics of neutron star superfluidity unfolds on a hierarchy 
of three distinct length-scales. The separation of these scales 
is useful, as often the physics of 
a neighboring scale enters a theory at a given scale in the
form of phenomenological constants, which can  potentially be 
fixed by comparison with measured observables.

At the {\it microscopic} level the physical scale of the order of 
fermi (fm$=10^{-13}$ cm) is set  by the nuclear 
forces. The long-range attractive interaction between nucleons leads 
to an instability of the normal state of the nuclear matter against 
Cooper pairing, in analogy to the microscopic Bardeen-Cooper-Schrieffer 
theory of superconductivity of metals. The nuclear 
forces control the size of the ``elementary bosons'' of the theory 
- the Cooper pairs, which appear as  weakly bound states of two fermions
near the Fermi surface.
The size of a Cooper pair, the {\it coherence length} $\xi$, 
is of the order of 10 fm. It sets, obviously, the lower scale on which 
the hydrodynamic description of superfluids breaks down. 
On length scales larger than $\xi$, the condensate of 
Cooper-pairs can be described by a single wave function
$\psi({\bf x})$, i.e., the condensate forms a macroscopically coherent state.

At the {\it local hydrodynamic} level the relevant physical
scale is set by the size of vortices -  macroscopic quantum objects,
whose fundamental property is the 
quantization of the circulation around a path encircling their 
core. The circulation is quantized in units 
of $2\pi\hbar$ since the condensate wave-function must be single 
valued function at each point of the condensate.
On writing 
$
\psi = \psi_0 e^{i\chi}, 
$
the gauge invariant superfluid velocities 
can be expressed through the gradient of
the phase of superfluid order parameter $\chi$ and the
vector potential, ${\bf A}$
\begin{equation}\label{eq:ASv}
{\bf v}_{\tau}=\frac{\hbar}{2m_{\tau}}{\bf \nabla }
\chi _{\tau}-\frac{e_{\tau}}{m_{\tau}c} {\bf A},
\end{equation}
where $ e_{\tau} \equiv (e, 0 )$ is the electric charge of
protons and neutrons respectively, $m_{\tau}$ is their bare mass;
$\tau \in \{n, p\}$, where $n$ and $p$ stand for neutrons and 
protons respectively.
Applying the curl operator to Eq. (\ref{eq:ASv}) 
and implementing quantization 
of the circulation (the phase of the superfluid
order parameter changes by $2\pi$ around a closed path) we find
\begin{eqnarray}\label{eq:AS:curlv}
\mathop{\rm curl} {\bf v_{\tau}} = \frac{\pi \hbar}{m_{\tau}}
{\bf  \nu}_{\tau} \sum_{j} \delta^{(2)} ({{\bf x}}-{{\bf x}}_{\tau j})
- \frac{ e_{\tau}}{m_{\tau}c}{\bf B} \equiv {\bf  \omega}_{\tau},
\end{eqnarray}
where $\pi \hbar/m_{\tau}$ is the circulation quantum, ${\bf  \nu}_{\tau}
\equiv {\bf  \omega}_{\tau}/\omega_{\tau}$ is a unit vector
along the vortex lines, ${{\bf x}}_{\tau j}$ defines the position
of a vortex line in the plane  orthogonal to the vector
${\bf  \nu}_{\tau}$, $\delta^{(2)}$ is a two--dimensional delta
function in this plane and ${\bf B}=\mathop{\rm curl} \, {\bf A}$ is the magnetic 
field induction.
The $j$--summation is over the sites of vortex lines. Note that
Eq. (\ref{eq:AS:curlv})  treats the vortex cores
as singularities in the plane orthogonal 
to $ \nu_{\tau}$, which is justified on scales larger 
than the coherence length of a condensate. The crossover from the 
local hydrodynamic scale to the microscopic scale can be studied within 
the Ginzburg-Landau theory as we briefly discuss below.

For a single vortex the integral of Eq. (\ref{eq:AS:curlv}) completely 
determines the superfluid pattern; as this equation is linear, for 
larger number of vortices the superfluid pattern is a superposition 
of the flows induced by each vortex. The resulting net
flow, obviously, depends 
on the arrangement of the vortices. It turns out that the integral 
of Eq. (\ref{eq:AS:curlv}) on the local hydrodynamic scale is
radically different from the one on the scales involving larger number of
vortices in a rotating superfluid.

To appreciate the difference in superfluid patterns on different 
scales let us look for vortex solutions in a neutral condensate 
in cylindrical geometry. The condensate 
wave function has the form
$\psi({\bf x}) =  f(r) e^{i\theta}$ in the polar-cylindrical coordinates
$(r,\theta, z)$; the neutron superfluid velocity upon 
integrating Eq. (\ref{eq:AS:curlv}) becomes
\begin{eqnarray}\label{eq:AS:vn}
{\bf v}_n = \frac{\hbar}{2m_nr}\hat \theta .
\end{eqnarray} 
The divergence of the superfluid velocity when $r\to 0$ is avoided
by introducing cut-off on scales of the order of the coherence length. 
The cut-off scale, for Fermi-liquids, can be understood by noting that  
an increase of  ${\bf v}_n$  when  $r\to 0$ causes an increase 
of the kinetic energy of Cooper pairs which
eventually becomes larger than the binding energy 
of a pair. The broken pairs will perform a rigid-body 
rotation with an angular velocity which scales 
as $Cr$ and is  regular in the
$r\to 0$ limit (here $C$ is a constant). 
This crossover can be seen from the 
well-known solutions of the Ginzburg-Landau equation
for the amplitude $f(r)$: 
\begin{eqnarray}\label{eq:AS:nondim}
\frac{d^2f}{d\zeta^2} + \frac{1}{\zeta}\frac{df}{d\zeta} - 
\frac{1}{\zeta^2}f + f -f^3 = 0,
\end{eqnarray}
where $\zeta = r/\xi_n$ and  $\xi_n$ is the size of the vortex
core. The asymptotic solutions of Eq. (\ref{eq:AS:nondim}) are
\begin{eqnarray} 
f(\zeta)= \left\{
\begin{array}{cc} C\zeta & \zeta \ll 1,\\
1 - (2\zeta^2)^{-1} & \zeta \to \infty ,\\
\end{array} \right.
\end{eqnarray}
while numerical solutions for  intermediate values of $\zeta$ 
show that  the condensate wave function 
is at half of its value in a homogeneous condensate when $\zeta = 1$. 
Note the long-range nature of the superfluid vortex velocity, and the 
resulting slow fall-off of the density perturbation in the condensate.
This behavior is specific to neutral condensates; for charged
condensates the super-current is screened exponentially on 
length scales of the order of the penetration depth $\lambda$. The solution 
of Eq. (\ref{eq:AS:curlv}) for a charged condensate is  
\begin{eqnarray}\label{eq:AS:vp}
{\bf v}_p = \frac{\hbar}{2m_p\lambda}
K_1\left(\frac{r}{\lambda}\right)\hat \theta ,
\end{eqnarray} 
where $K_1$ 
is the Bessel function of imaginary argument; as for $r\gg \lambda$,
$K_1(r/\lambda) \simeq {\rm exp} (-r/\lambda)$, therefore the superfluid 
circulation decays exponentially; in the opposite 
limit $r\ll \lambda$, $K_1(r/\lambda) \simeq \lambda/r$ and 
Eq. (\ref{eq:AS:vp}) assumes a form identical to Eq. (\ref{eq:AS:vn}).

At the {\it global hydrodynamic} level the relevant scales
are of the order of the size of the system, which in 
neutron stars is of the order of kilometers. 
On these scales the hydrodynamic and thermodynamic
variables are course-grained quantities, 
i.e. they are averages over a large number of vortices. 
The solution (\ref{eq:AS:vn}) does not minimize the energy 
$E-{\bf L\cdot \Omega}$ 
of a rotating superfluid, where $E$ and $\bf L$ represent
the kinetic energy and  the angular momentum respectively.
The energy acquires its minimum for a superfluid flow which 
to a high precision mimics a rigid body rotation i.e., 
$v_n = \Omega r$, where $r$ is the  distance from the rotation axis;
(small deviation occur only at the bounding surface of the 
superfluid).
On the global hydrodynamic scales a transition 
to a continuum vortex distribution can be carried out on 
the right-hand side of  
Eq. (\ref{eq:AS:curlv}) by defining vortex densities $n_{\tau}=
\sum_{j} \delta^{(2)} ({\bf x}-{\bf x}_{\tau j})$. 
Since for rigid-body rotations the curl of ${\bf v}_n$ is simply 
$2\Omega$, the number density of vortices in the neutron superfluid 
is related to the macroscopic  angular velocity of 
the neutron condensate by the familiar Feynman formula 
\begin{eqnarray}\label{eq:AS:nn}
n_n = \frac{2m_n\Omega}{\pi\hbar}.
\end{eqnarray}
For typical pulsar periods, $0.05 < P < 0.5$ s,  
$n_n\simeq 6.3.\times 10^3 \, P^{-1}\sim 10^4$-$10^5$ per cm$^2$.
The minute difference between the superfluid and normal 
angular velocities in a neutron star decelerating under external 
breaking torques is neglected here. 
For a charged superfluid Eq. (\ref{eq:AS:curlv}) can be transformed 
to a contour integral over a path where ${\bf v}_p=0$, 
as the super-current is screened beyond 
the magnetic field penetration depth $\lambda$.
Again, going over to the continuum vortex limit we find
\begin{eqnarray}\label{eq:AS:np}
n_p =\frac{B}{\Phi_0}\simeq  5\times 10^{18}
{~~\rm cm}^{-2}.
\end{eqnarray}
Note that the number of proton vortices per neutron vortex is
$n_p/n_n\sim 10^{13}-10^{14}$ independent of their arrangement.
The energy of a bundle of neutron or proton vortices is 
minimized by a triangular lattice with a unit cell area 
$$
n_{\tau}^{-1}=(\sqrt{3}/2)~d_{\tau}^2.
$$
The length of a ``basis vector'' of such a lattice in a neutron 
condensate (the neutron inter-vortex distance) is 
\begin{eqnarray}
   d_n= \left(\frac{\pi\hbar}{\sqrt{3}\, m_n\, \Omega}   \right)^{1/2}.
\end{eqnarray}
For the inter-vortex distance in the  proton condensate we find
\begin{eqnarray}
  d_p=\left(\frac{2\, \Phi_0}{\sqrt{3} \, B}   \right)^{1/2},
\end{eqnarray}
where $B$ is the mean magnetic field induction.
Using the estimates given in Eqs. (\ref{eq:AS:nn}) and (\ref{eq:AS:np}) 
we find that the neutron and proton inter-vortex distances are
$d_n \sim 10^{-2}-10^{-3}$ cm and $d_p\sim 10^{-9}$ cm respectively.
For typical values of the microscopic parameters  
the penetration depth is of the order of 100 fm, therefore the conditions
$\xi_n\ll d_n$  and $\xi_p\ll {\rm min }( \lambda, ~d_p)$ are satisfied
and  the use of the hydrodynamics on the local scale is valid. It is 
also clear that the global hydrodynamics can be applied on  the scales
that are much larger than $d_n$ (a fraction of millimeter).  

The remainder of this review 
concentrates on the physics of the global hydrodynamic 
scale and on neutral superfluids only, as the dominant fraction of the 
moment of inertia of a neutron star resides in the neutron fluid and it plays 
the main role in the hydrodynamic oscillations of the star. 
Charged superfluids will be absorbed in the normal fluid of 
the star formally, as they are coupled to the electron liquid 
via electro-magnetic forces on short time-scales. Note that 
their role is crucial in controlling 
the mutual friction on the local hydrodynamic
scale; however the physics of this scale will 
enter the theory on the global scale 
via phenomenological constants, which we will treat as 
free parameters of the theory.

\section{Two-fluid Newtonian superfluid hydrodynamics}

The superfluid phases in neutron stars coexist with normal fluids whose
interaction with superfluid vortices leads to the effect
of the {\it mutual friction} between a superfluid and normal fluid. 
A phenomenological description of this effect is based on the
two-fluid dissipative hydrodynamics. A particularly simple and 
transparent formulation which, with some care, can be taken 
over to describe superfluids in neutron stars, was developed
by Bekarevich and Khalatnikov for liquid He$^4$ \cite{BK}. 
It is interesting that the general 
form of mutual friction forces can be obtained by utilizing only
the conservation laws and some reasonable assumptions on the form 
of the dissipation. Although the  superfluids in a neutron star can 
be assumed to be at zero temperature,
i.e., the number of quasi-particle excitations 
is small, they coexist with a normal liquid of electrons in the
core and a nuclear lattice in the crusts. Hence, entropy is irreversibly 
produced due to various dissipative mechanism in the normal fluid
even though the superfluid matter is effectively at zero temperature. 
Formally, the electron liquid and the nuclear lattice in the crusts 
take over the role of quasi-particle excitations 
in the superfluid hydrodynamics of liquid He$^4$.

The conservation of the combined mass of the two fluids is given 
by the continuity equation
\begin{eqnarray} 
\frac{\partial\rho}{\partial t} + \nabla\cdot J &=&0,
\label{eq:AS2.2}
\end{eqnarray}
where the net mass  $\rho= \rho_S +\rho_N$
is the sum of the masses of  superfluid and normal fluid,
$ J = \rho_N {\bf v}_N + \rho_S{\bf v}_S$ is the mass current
(hereafter the indexes $S$ and $N$ refer to superfluid and 
normal fluid, respectively.) 
The total momentum conservation is 
\begin{eqnarray}
\frac{\partial J_i}{\partial t} + \frac{\partial
P_{ik}}{\partial x_k} &=& 0, 
\end{eqnarray}
where $P_{ik}$ is the stress energy tensor. 
The time evolution of the entropy, $S$, of normal fluid 
can be written as
\begin{eqnarray} 
\frac{\partial S}{\partial t}  +  \nabla \cdot S {\bf v}_n = \frac{R}{T},
\end{eqnarray}
where $R$ is the dissipative function, $T$ is the temperature; finally,
the conservation of the energy, $E$, reads
\begin{eqnarray} 
\frac{\partial E}{\partial t}  +  \nabla \cdot  Q = 0 ,
\end{eqnarray}
where $ Q$ is the energy current.
Equations above should be supplemented by the Euler equations 
for the superfluid and the  normal fluid
\begin{eqnarray}\label{eq:AS:EULER_S}
\rho_S\left[ \frac{\partial {\bf v}_S}{\partial t} 
 + ({\bf v}_S\cdot
\nabla)\cdot {\bf v}_S\right]& = & -
\frac{\rho_S}{\rho}\nabla p -\rho_S\nabla\phi + F,\\
\label{eq:AS:EULER_N}
\rho_N\left[\frac{\partial {\bf v}_N }{\partial t} 
+ ({\bf v}_N\cdot\nabla)\cdot {\bf v}_N \right]&=&-\frac{\rho_N}{\rho}\nabla p
- \rho_N\nabla \phi +\eta_N\Delta {\bf v}_N -  F,
\end{eqnarray}
where $ F$ is the mutual friction force, and $\eta_N$ is the 
viscosity of the normal fluid, $\phi$ is the Newtonian 
gravitational potential. To determine the unknowns in the 
hydrodynamic equations, let us write the total energy of the fluid 
in the frame in which the normal fluid is at rest as
\begin{eqnarray}\label{eq:ASAS_ENERGY}
E =\frac{1}{2}\rho v_S^2 
+ ( J -  \rho {\bf v}_S) \cdot {\bf v}_S + {\cal E},
\end{eqnarray}
where the internal energy ${\cal E}$
is given by the second law of thermodynamics as
\begin{eqnarray} 
d{\cal E} = TdS + \mu d\rho 
+ ({\bf v}_N-{\bf v}_S) \cdot d( J -  \rho {\bf v}_S) 
+ \Lambda d\omega .
\end{eqnarray}
The energy due to the vorticity is represented by the term  
which is proportional to $\omega =  \nabla \times {\bf v}_S$. 
Differentiating Eq. (\ref{eq:ASAS_ENERGY}) with respect to time 
and eliminating the time derivatives using the conservation laws
above we recover the conservation of the energy 
\begin{eqnarray}\label{eq:ASAS_RESOLVED} 
\frac{\partial E}{\partial t} +  \nabla \cdot    Q &=& 
 R + \left(P'_{ik} - \Lambda \omega \delta_{ik} 
+ \Lambda\frac{\omega_i \omega_k}{\omega}\right)
\frac{\partial v_{Ni} }{\partial x_k}\nonumber\\
&+&\left( J -\rho{\bf v}_N+ \nabla \times \Lambda  \nu 
\right)\cdot \left\{ F + \left[({\bf v}_S - {\bf v}_N) 
\times  \omega \right] \right\} = 0,
\end{eqnarray}
where $P'_{ik}$ is the part of the stress tensor associated with the 
vorticity; the explicit form of the energy current is not 
indicated since it will not be used in the following. Since the dissipative 
function $R$ must be positive, the remaining terms on the right-hand side 
of Eq. (\ref{eq:ASAS_RESOLVED}) must be quadratic forms for small deviations
from  equilibrium. This implies that the most general 
form of the mutual friction force is
\begin{eqnarray}
\label{eq:AS2.4}
 F &=& -\left[\omega\times\left(\nabla\times
\Lambda\nu\right) \right]- \beta\left[\,\nu\times
\left[\,\omega\times({\bf v}_N-{\bf v}_S-
\nabla\times\Lambda\nu)\,\right]\,\right]\nonumber \\
&-&\beta'\left[\,\omega\times ({\bf v}_N -{\bf v}_S
- \nabla \times\Lambda\nu)\right] 
+\beta''\nu\cdot\left[\, \omega\cdot ({\bf v}_N -{\bf v}_S
- \nabla \times \Lambda\nu)\right],
\end{eqnarray}
where $\beta$,  $\beta'$ and  $\beta''$ 
are phenomenological coefficients. On substituting the mutual 
friction force in the Euler equation for the superfluid, Eq. 
(\ref{eq:AS:EULER_S}), we see that the vorticity propagates 
with a velocity ${\bf v}_L$, that is  
\begin{eqnarray} 
\frac{\partial {\omega}}{\partial t}
={\nabla}\times
\left({\bf v}_L\times{\omega}\right),
\end{eqnarray}
which is defined, assuming $\beta''\ll \beta\, , \beta'$, as 
\begin{eqnarray}
{\bf v}_L &=&{\bf v}_S +\nabla\times\Lambda\nu  
+\beta'({\bf v}_N -{\bf v}_S- \nabla \times\Lambda\nu) \nonumber\\ 
&+& \beta\left[\,\omega\times({\bf v}_N-{\bf v}_S-
\nabla\times\Lambda\nu)\,\right].
\end{eqnarray}
The latter equation can be put in a form reflecting the
balance of forces acting on a vortex
\begin{eqnarray}\label{eq:AS2.4.2}
\rho_S\left[\left({\bf v}_S
+\nabla\times\Lambda\nu-{\bf v}_L\right)\times
\omega\right] - \eta\left({\bf v}_L-{\bf v}_N\right)
+\eta'\left[\left({\bf v}_L-{\bf v}_N\right)\times\nu\right] = 0,
\end{eqnarray} 
with the new phenomenological coefficients $\eta$ and $\eta'$ defined as 
\begin{eqnarray}\label{eq:AS:etas}
\beta = \frac{\eta\rho_S\omega}{\eta^2+\left(\rho_S\omega-\eta'\right)^2},\quad
\beta' = 1- 
\frac{\rho_S\omega\,(\rho_S\omega-\eta')}{\eta^2+\left(\rho_S\omega-\eta'\right)^2}.
\end{eqnarray}
The first term in Eq. (\ref{eq:AS2.4.2}) is a
non-dissipative lifting force due to a superflow past the vortex
(the Magnus force).
The remaining terms reflect the friction between the vortex 
and the normal fluid. The coefficients $\eta$ and  $\eta'$,
therefore,  measure the 
friction parallel and orthogonal to the vortex motion in 
the plane orthogonal to the average direction 
of the vorticity. 
A nonzero $\beta''$ implies friction along the {\it average}
direction of the vorticity, which is possible if 
vortices are oscillating, or are subject to other  
deformations in the plane orthogonal to the rotation. 
One may assume, at least under stationary conditions, 
that $\beta''\ll \beta, \beta'$.

In the both limits of either {\it strong coupling} 
($\eta\gg \rho_S\omega$) or {\it weak 
coupling} ($\eta\ll \rho_S\omega$) between a vortex and the normal 
fluid, one finds that $\beta\to 0$ as a function of $\eta$, 
with the maximum $\beta_{\rm max} = 0.5$ at 
$\eta= \rho_S\omega$. In the strong coupling limit 
$\beta'(\eta)\to 1$, while in the opposite weak coupling 
limit  $\beta'(\eta)\to 0$
( generally we assume that the quasi-particle--vortex 
scattering kinematics implies $\eta'\ll \eta$ 
and that for the relevant densities $\eta'\ll\rho_S\omega$).
 Note that $\beta'(\eta)$ approaches its asymptotic 
strong-coupling values quadratically, while $\beta(\eta)$ does so linearly; 
the asymptotic behavior for large $\eta$'s, 
therefore, is dominated by $\beta(\eta)$. 

\section{Virial equations and their perturbations}

Virial equations of various order are constructed by taking 
moments of the hydrodynamic equations. Since the computation of the 
perturbations of the these virial equations is central 
to the theory of superfluid ellipsoids we review this 
somewhat technical issue in this section.
The reader who is interested only in the physics of superfluid 
oscillations can proceed to the next section where we discuss 
the oscillations of Maclaurin spheroids.

The equations of motion (\ref{eq:AS:EULER_S}) and (\ref{eq:AS:EULER_N}),
written in a frame rotating with angular velocity
${\mbox{\boldmath $\omega$}}$ relative to some inertial 
coordinate reference system, can be combined in a single equation

\begin{eqnarray}
\rho_\alpha\left({\partial\over\partial t}+u_{\alpha,j}
{\partial\over\partial x_j}\right)  u_{\alpha,i}
&=&-{\partial p_\alpha\over
\partial x_i}-\rho_\alpha{\partial\phi\over\partial x_i}
+{1\over 2}\rho_\alpha{\partial\vert{\mbox{\boldmath $\omega$}}
{\mbox{\boldmath $\times$}}{\bf x}\vert^2
\over\partial x_i}\nonumber\\
&+&2\rho_\alpha\epsilon_{ilm}u_{\alpha,l}\Omega_m +F_{\alpha\beta,i},
\label{eq:AS:euler}
\end{eqnarray}
where the subscript $\alpha\in\{S,N\}$ identifies the fluid component,
and Latin subscripts denote coordinate directions; $\rho_\alpha$,
$p_\alpha$, and ${\bf u}_\alpha$ are the density, pressure, and
velocity of fluid $\alpha$.
The two fluids are coupled to one
another  by  mutual gravitational attraction and the mutual friction 
force ${\bf F}_{\alpha\beta}$ [Eq. (\ref{eq:AS2.4})].
The gravitational potential $\phi$ is derived from
\begin{eqnarray}
\nabla^2\phi=\nabla^2(\phi_S+\phi_N)=4\pi G[\rho_S({\bf x})+\rho_N({\bf x})];
\end{eqnarray}
the individual fluid potentials $\phi_\alpha$ obey
$\nabla^2\phi_\alpha=4\pi G\rho_\alpha$.
For a normal-superfluid mixture the mutual friction force,
written in components, is
\begin{eqnarray}
F_{SN,i}=-\rho_S\omega_S\beta_{ij}(u_{S,j}-u_{N,j}),
\end{eqnarray}
where the mutual friction tensor is
\begin{eqnarray}
\beta_{ij}=\beta\delta_{ij}+\beta^\prime\epsilon_{ijm}\nu_m
+(\beta^{\prime\prime}-\beta)\nu_i\nu_j,
\label{eq:AS:betaijdef}
\end{eqnarray}
with $\beta$, $\beta^{\prime}$
 and $\beta^{\prime\prime}$ being the mutual friction
coefficients, and ${\mbox{\boldmath $\omega$}}_S
={\mbox{\boldmath $\nu$}}\omega_S\equiv{\mbox{\boldmath $\nabla\times$}}{\bf u}_S$.

The Euler equation (\ref{eq:AS:euler}) can be extended to include
external gravitational sources, for example 
the tidal potential of an external point source 
of gravity acting on an ellipsoid (Roche ellipsoids). 
We will not discuss here the stability and oscillations of 
superfluid counterparts of the  classical Roche binaries; their 
relative oscillation modes are derived in Ref. \cite{SW2000}.  

The general strategy for finding the equilibrium shapes
of ellipsoidal figures and modes of their oscillations within the tensor 
virial method consist of (i) constructing moments of the 
hydrodynamic equations describing fluid motions in the 
rotating frame; (ii) computing Eulerian perturbations
of the resulting virial equations; (iii) expressing these 
perturbations in terms of the virials of various order; these
are defined as the moments of the Lagrangian displacement 
${\mbox{\boldmath $\xi$}}_{\alpha,i}$ of fluid $\alpha$:  
\begin{eqnarray}
 V_{\alpha, i} &\equiv& \int_{V_{\alpha}}d^3 x\,\rho\, 
{\mbox{\boldmath $\xi$}}_{\alpha,i},
\quad ({\rm first ~order}) \\
 V_{\alpha, i;j} &\equiv& \int_{V_{\alpha}}d^3 x\,\rho\, 
{\mbox{\boldmath $\xi$}}_{\alpha,i} x_j,
\quad ({\rm second ~order}) \\
 V_{\alpha, i;jk} &\equiv& \int_{V_{\alpha}}d^3 x\,\rho\, 
{\mbox{\boldmath $\xi$}}_{\alpha,i} x_{j}x_{k},
\quad ({\rm third ~order}) \\
\dots\nonumber
\end{eqnarray}
The advantage of using the homogeneous ellipsoidal approximation 
is that the perturbations of the gravitational energy tensor of an ellipsoid 
can be expressed in terms of the {\it index symbols} defined 
as (cf. EFE Chap. 3)
\begin{eqnarray} 
A_{ijk\dots}& =& a_1a_2a_3\int_0^{\infty}
\frac{du}{\Delta(a_i^2+u)(a_j^2+u)(a_k^2+u)\dots},\\
B_{ijk\dots}& =& a_1a_2a_3\int_0^{\infty}
\frac{udu}{\Delta(a_i^2+u)(a_j^2+u)(a_k^2+u)\dots},
\end{eqnarray} 
where $\Delta^2 = (a_1^2+u)(a_2^2+u)(a_3^2+u)$ and $a_1$, 
$a_2$, and $a_3$ are the semi-axis of the ellipsoid. This 
strategy is described in detail in EFE. Below, 
we concentrate on its extension to superfluids
with an  emphasis on the new effects of 
mutual friction and mutual gravitational 
attraction of the superfluid and normal fluid.

\subsection{First order virial equations}

On taking the zeroth moment 
 of Eq. (\ref{eq:AS:euler}) which amounts to integrating over 
$V_\alpha$ we obtain the ``first order `virial' equation''\footnote{The 
word ``virial'' is in quotes because the equations are intrinsically
dissipative.} 
\begin{eqnarray}
{d\over dt}\biggl(\int_{V_\alpha}{d^3x\rho_\alpha u_{\alpha,i}}
\biggr)&=&2\epsilon_{ilm}\Omega_m\int_{V_\alpha}{d^3x\rho_\alpha
u_{\alpha,l}}+(\Omega^2\delta_{ij}-\Omega_i\Omega_j)
\int_{V_\alpha}{d^3x\rho_\alpha x_j}
\nonumber\\& & 
-(1-\delta_{\alpha\beta})
\int_{V_\alpha}{d^3x\rho_\alpha{\partial\phi_\beta\over\partial x_i}}
+\int_{V_\alpha}{d^3xF_{\alpha\beta,i}}.
\label{eq:AS:v1}
\end{eqnarray}
Note that we impose the boundary condition $p_\alpha=0$ for each fluid 
 on the bounding surface of $V_\alpha$. The fluids 
are not restricted to occupy the same volume.
Apart from simple doubling of the number of the 
inertial forces, which do not couple the two fluids, 
there are two forces that do couple them: gravity and friction.
The net mutual gravitational force between the fluids vanishes only
if they (i)~occupy the same volume and (ii)~have densities
that are proportional to one another (i.e. $\rho_S\propto
\rho_N$). The mutual friction force is nonzero as long as the fluids
move relative to one another. 
Although the mutual friction force is nonzero only in
the {\it overlap} volume of the two fluids - a  restriction which is
necessary to derive conservation of total momentum for the combined
fluids - it would be effective throughout the entire volume of
fluids because the force is mediated by a macroscopically extended
vortex lattice.

For isolated single-fluid ellipsoids the first harmonic oscillations
are trivial, since they correspond to the motions of the center-of-mass
of an ellipsoid and can be eliminated by a transformation to the reference
frame whose origin is located at  
the center-of-mass of the ellipsoid. For two-fluid ellipsoids the fluid
motions include  the counter-phase (relative) oscillations of two fluids, 
which can not be eliminated by any transformation.
These are the only 
new type of oscillations for the superfluid Maclaurin and Jacobi 
ellipsoids (the solitary ellipsoids with vanishing internal motions). 
In the case of ellipsoids in an external gravitational field (e.g.
the Roche ellipsoids), 
the CM motions are not trivial 
any more, since the external (inhomogeneous by assumption) source of 
gravitational field breaks the translational symmetry of the problem. Hence, 
apart from the relative oscillations of two fluids, the CM
oscillations become non-trivial. 

Consider first the variation of the first order 
virial equation under the influence
of perturbations. The variations of the inertial terms 
proceeds in full analogy to EFE. Here we 
concentrate on variations of the new terms corresponding to the 
mutual gravitational attraction and mutual friction.
The variation of the first force is 
\begin{eqnarray}
-\delta\int_{V_\alpha}{d^3x\rho_\alpha{\partial\phi_\beta
\over\partial x_i}}&=&
-\int_{V_\alpha}{d^3x\rho_\alpha({\bf x})
\xi_{\alpha,l}({\bf x}){\partial\over\partial x_l}}
\int_{V_\beta}{d^3x^{\prime}\rho_\beta({\bf x}^{\prime})(x_i-x^{\prime}_i)
\over\vert{\bf x}-{\bf x}^{\prime}\vert^3}
\nonumber\\
&+&\int_{V_\beta}{d^3x\rho_\beta({\bf x})\xi_{\beta,l}({\bf x}){\partial\over\partial x_l}}
\int_{V_\alpha}{d^3x^{\prime}\rho_\alpha({\bf x}^{\prime})(x_i-x^{\prime}_i)
\over\vert{\bf x}-{\bf x}^{\prime}\vert^3}
\label{eq:ASgravab}
\end{eqnarray}
which is manifestly antisymmetric on $\alpha\leftrightarrow\beta$.
Assuming $V_\alpha=V_\beta=V$ 
and $\rho_\alpha=f_\alpha\rho({\bf x})$ in the background
equilibrium, we can simplify this to
\begin{eqnarray}
-\delta\int_{V_\alpha}
d^3x\rho_\alpha\frac{\partial\phi_\beta}
{\partial x_i}&=&f_\alpha f_\beta
\int_V d^3x\rho({\bf x})[\xi_{\beta,l}
({\bf x})-\xi_{\alpha,l}({\bf x})]\nonumber\\
&\times&\frac{\partial}{\partial x_l}
\int_V\frac{d^2x^{\prime}\rho({\bf x}^{\prime})(x_i-x^{\prime}_i)}
{\vert{\bf x}-{\bf x}^{\prime}\vert^3}.
\end{eqnarray}
Although we simplified the final answer by assuming that the fluids 
occupy identical volumes and have proportional densities 
in the background state, we could not have derived the correct 
perturbation of the first order virial theorem if we had not allowed 
the volumes to differ. 

For uniform ellipsoids, we can simplify the mutual 
gravitational term further. First, we note that the second integral
is simply the derivative of the gravitational 
potential  which at any interior point of 
a homogeneous ellipsoid is
\begin{eqnarray}\label{eq:AS:ellips_phi}
\phi({\bf x})=-\pi G\rho\biggl(I-\sum_{k=1}^3A_kx_k^2\biggr),
\end{eqnarray}
where $I$ is a constant.
Thus, the mutual gravitational contribution to the equation 
of motion for the perturbed center-of-mass is
\begin{eqnarray}
2\pi G\rho^2A_if_\alpha f_\beta 
\int_V{d^3x(\xi_{\beta,i}-\xi_{\alpha,i})}.
\end{eqnarray}
If in  the background state, the two fluids move 
together or are stationary, the variation of the mutual friction 
force becomes
\begin{eqnarray}
\delta\int_V{d^3x F_{\alpha\beta,i}}=-{\cal S}_{\alpha\beta}
{d\over dt}\biggl[f_S\int_V{d^3x\rho({\bf x})\omega_S\beta_{ij}
(\xi_{S,j}-\xi_{N,j})}\biggr],
\end{eqnarray}
where ${\cal S}_{\alpha\beta}=0$ if $\alpha=\beta$, 1 if $\alpha=S$ and
$\beta=N$, and $-1$ if $\alpha=N$ and $\beta=S$. Collecting terms
we find the first order virial equation
\begin{eqnarray}
f_{\alpha} \frac{d^2 V_{\alpha,i} }{dt^2}&=&
2\epsilon_{ilm}f_{\alpha}\Omega_m\frac{d}{dt} V_{\alpha,l}
+(\Omega^2\delta_{ij} -\Omega_i\Omega_j) f_{\alpha} V_{\alpha,j}\nonumber\\
&-&2\pi G \rho A_i f_{\alpha} f_{\beta} \left(V_{\alpha,i}-
V_{\beta,i}\right)
-{\cal S}_{\alpha\beta}f_S\omega_S \beta_{ij}
\left(V_{\alpha,j}-V_{\beta,j}\right).
\label{eq:AS:PERTURB}
\end{eqnarray}
The CM and relative motions can be decoupled by defining
\begin{eqnarray}
V_{i}\equiv f_SV_{S,i}+f_NV_{N,i},\qquad
U_{i}\equiv V_{S,i}-V_{N,i}.
\label{eq:ASNEW_VIRIALS}
\end{eqnarray}
The CM motions of two fluids are trivial
(as they can be eliminated by a linear transformation of the
reference frame) and, therefore,  $V_i =0$. The virial equation
describing the relative motions of the two fluids is
\begin{eqnarray}
\frac{d^2}{dt^2}U_i = 2 \epsilon_{ilm}\Omega_m\frac{d}{dt} U_l
+(\Omega^2\delta_{ij}-\Omega_i\Omega_j) U_j - 2 A_i U_i
-2 \Omega \tilde \beta_{ij} \frac{d}{dt}U_j.
\label{eq:AS:monopole}
\end{eqnarray}
From the latter equation it is straightforward to compute
the first harmonic relative oscillation modes of 
irrotational ellipsoids, which is done in the next section
for Maclaurin spheroids.

\subsection{Second order virial equations}

Taking the first moment of Eq. (\ref{eq:AS:euler}) results in the
second order `virial' equation
\begin{eqnarray}
{d\over dt}\biggl(\int_{V_\alpha}{d^3x\rho_\alpha x_ju_{\alpha,i}}
\biggr)&=&2\epsilon_{ilm}\Omega_m\biggl(\int_{V_\alpha}
{d^3x\rho_\alpha x_ju_{\alpha,l}}\biggr)
+\Omega^2I_{\alpha,ij}-\Omega_i\Omega_kI_{\alpha,kj}
\nonumber\\
&+&2{\cal T}_{\alpha,ij}+\delta_{ij}\Pi_{\alpha}
+{\cal M}_{\alpha,ij}+(1-\delta_{\alpha\beta}){\cal M}_{\alpha\beta,ij}
+{\cal F}_{\alpha\beta,ij},\nonumber\\
\label{eq:AS:v2}
\end{eqnarray}
where
\begin{eqnarray}
I_{\alpha,ij}&\equiv&\int_{V_\alpha}{d^3x\,\rho_\alpha x_ix_j}\\
\Pi_\alpha&\equiv&\int_{V_\alpha}{d^3x\,p_\alpha}\\
{\cal T}_{\alpha,ij}&\equiv&{1\over 2}\int_{V_\alpha}{d^3x\rho_\alpha
u_{\alpha,i}u_{\alpha,j}}\\
{\cal M}_{\alpha,ij}&\equiv&
-{G\over 2}\int_{V_\alpha}{d^3x\,d^3x^{\prime}
\rho_\alpha({\bf x})\rho_\alpha({\bf x}^{\prime})
(x_i-x^{\prime}_i)(x_j-x^{\prime}_j)\over\vert{\bf x}-{\bf x}^{\prime}\vert^3}\\
{\cal M}_{\alpha\beta,ij}&\equiv&
-G\int_{V_\alpha}{d^3x}\int_{V_\beta}{d^3x^{\prime}\rho_\alpha({\bf x})
\rho_\beta({\bf x}^{\prime})x_j(x_i-x^{\prime}_i)\over\vert{\bf x}-{\bf x}^{\prime}\vert^3}
\\
{\cal F}_{\alpha\beta,ij}&\equiv&\int_{V_\alpha}{d^3x\,x_j
F_{\alpha\beta,i}}.
\end{eqnarray}
When there is only one fluid present, this equation reduces 
to the one found in Chap. 2 of EFE.
Again we consider only variations of the new terms 
in the second order virial equation due to the mutual gravitational 
attraction (${\cal M}_{\alpha\beta,ij}$)
and mutual friction (${\cal F}_{\alpha\beta,ij}$). 
The first variation is
\begin{eqnarray}
\delta{\cal M}_{\alpha\beta,ij}&=&
-Gf_\alpha f_\beta\biggl\{\int_V{d^3x\rho({\bf x})\xi_\alpha({\bf x}){\partial\over
\partial x_l}}\int_V{d^3x^{\prime}\rho({\bf x}^{\prime})(x_i-x^{\prime}_i)(x_j-x^{\prime}_j)\over
\vert{\bf x}-{\bf x}^{\prime}\vert^3}\nonumber\\
& &+\int_V{d^3x\rho({\bf x})[\xi_{\alpha,l}({\bf x})-\xi_{\beta,l}({\bf x})]
{\partial\over\partial x_l}}
\int{d^3x^{\prime}\rho({\bf x}^{\prime})(x_i-x^{\prime}_i)x^{\prime}_j\over
\vert{\bf x}-{\bf x}^{\prime}\vert^3}\biggr\},
\end{eqnarray}
where we have specialized to backgrounds with proportional densities and
identical bounding volumes.
The first term in the brackets can be combined with 
$\delta{\cal M}_{\alpha,ij}$.
The resulting
equation can be written more compactly in terms of the functions
\begin{eqnarray}
{\cal B}_{ij}&\equiv& G\int_V{d^3x^{\prime}\rho({\bf x})(x_i-x^{\prime}_i)(x_j-x^{\prime}_j)\over
\vert{\bf x}-{\bf x}^{\prime}\vert^3},\\
{\partial{\cal D}_j\over\partial x_i}&\equiv&
-G\int_V{d^3x^{\prime}\rho({\bf x}^{\prime})
x^{\prime}_j(x_i-x^{\prime}_i)\over\vert{\bf x}-{\bf x}^{\prime}\vert^3},
\label{eq:ASddef}
\end{eqnarray}
which are related by
\begin{eqnarray}
{\partial{\cal D}_j\over\partial x_i}
={\cal B}_{ij}-x_j{\partial\phi\over\partial x_i};
\end{eqnarray}
we find
\begin{eqnarray}
\delta{\cal M}_{\alpha,ij}+(1-\delta_{\alpha\beta})\delta{\cal M}_{\alpha\beta,ij}
&=&-f_\alpha\int_V{d^3x\rho\xi_{\alpha,l}{\partial{\cal B}_{ij}\over\partial x_l}}
\nonumber\\
&+&f_\alpha f_\beta\int_V{d^3x\rho(\xi_{\alpha,l}-\xi_{\beta,l})
{\partial^2{\cal D}_j\over\partial x_l\partial x_i}}.
\end{eqnarray}
For the uniform ellipsoids (cf. EFE, Chap. 3, Eqs. [125] and [126]),
\begin{eqnarray}
{{\cal D}_j\over\pi G\rho}
&=&a_j^2x_j\biggl(A_j-\sum_{k=1}^3A_{jk}x_k^2\biggr),\\
{{\cal B}_{ij}\over\pi G\rho}
&=&
2B_{ij}x_ix_j
+a_i^2\delta_{ij}\biggl(A_i-\sum_{i=1}^3A_{ik}x_k^2\biggr),\\
\frac{1}{\pi G\rho}{\partial^2{\cal D}_j\over\partial x_l \partial x_i}
&=&
2B_{ij}(\delta_{il}x_j+\delta_{jl}x_i)
-2a_i^2\delta_{ij}A_{il}x_l.
\end{eqnarray}
Using these results, and defining symmetric in their indexes
second order virials as 
\begin{eqnarray}
V_{\alpha,ij}=V_{\alpha,i;j}+V_{\alpha,j;i},
\label{eq:AS:vdef}
\end{eqnarray}
we finally obtain
\begin{eqnarray}
{\delta{\cal M}_{\alpha,ij}+(1-\delta_{\alpha\beta})\delta{\cal M}_{\alpha\beta,ij}\over
\pi G\rho}=
-f_\alpha\biggl(2B_{ij}V_{\alpha,ij}-a_i^2\delta_{ij}\sum_{l=1}^3A_{il}V_{\alpha,ll}
\biggr)\nonumber\\
-a_j^2f_\alpha f_\beta\biggl[2A_{ij}(V_{\alpha,ij}-V_{\beta,ij})
+\delta_{ij}\sum_{l=1}^3A_{il}(V_{\alpha,ll}-V_{\beta,ll})\biggr].
\label{eq:AS:mutgrav}
\end{eqnarray}
For the perturbations of mutual friction force we find
\begin{eqnarray}
\delta\int_{V_\alpha}{d^3xx_jF_{\alpha\beta,i}}=
-{\cal S}_{\alpha\beta}f_S\int_V{d^3x\rho\omega_Sx_j\beta_{ik}\biggl(
{d\xi_{S,k}\over dt}-{d\xi_{N,k}\over dt}\biggr)}.
\label{eq:ASmfdef}
\end{eqnarray}
For perturbations of uniform ellipsoids, 
$\omega_S$ and $\rho$ are independent
of position in the unperturbed background, 
and we may also assume that $\beta_{ij}$
is constant; 
for backgrounds in which there 
are no fluid motions
\begin{eqnarray}
\delta\int_{V_\alpha}{d^3xx_jF_{\alpha\beta,i}}=-{\cal S}_{\alpha\beta}f_S\rho\omega_S
\beta_{ik}\biggl({dV_{\alpha,k;j}\over dt}-{dV_{\beta,k;j}\over dt}\biggr).
\end{eqnarray}
Thus the second order virial equation for a fluid $\alpha$,
in a frame rotating with an angular velocity $\Omega$, is
\begin{eqnarray}
f_{\alpha}{d^2V_{\alpha,i;j}
\over dt^2}&=&2\epsilon_{ilm}\Omega_mf_{\alpha}
{dV_{\alpha,l;j}\over dt}
+\Omega^2f_{\alpha}V_{\alpha,ij}
-\Omega_i\Omega_kf_{\alpha}V_{\alpha,kj}+\delta_{ij}\delta\Pi_{\alpha}
\nonumber\\
&-&f_{\alpha}\pi G\rho\biggl(2B_{ij}V_{\alpha,ij}
-a_i^2\delta_{ij}\sum_{l=1}^3A_{il}V_{\alpha,ll}
\biggr)
\nonumber\\
&-&a_j^2f_{\alpha} f_{\beta}\pi G\rho\biggl[2A_{ij}(V_{\alpha,ij}
-V_{\beta,ij})+\delta_{ij}
\sum_{l=1}^3A_{il}(V_{\alpha,ll}-V_{\beta,ll})\biggr]
\nonumber\\
&-&{\cal S}_{\alpha\beta}f_{\alpha}\omega_S
\beta_{ik}\frac{d}{dt}
\biggl(V_{\alpha,k;j}-V_{\beta,k;j}\biggr).
\label{eq:AS:2nd_order_virial}
\end{eqnarray}
We can replace these equations with a different set by defining
\begin{eqnarray}\label{eq:AS:VU}
V_{i;j}\equiv f_SV_{S,i;j}+f_NV_{N,i;j}\qquad
U_{i;j}\equiv V_{S,i;j}-V_{N,i;j}.
\end{eqnarray}
In terms of these new quantities we find 
\begin{eqnarray}\label{eq:AS:CM_virial}
{d^2V_{i;j}
\over dt^2}&=&2\epsilon_{ilm}\Omega_m
{dV_{l;j}\over dt}+\Omega^2V_{ij}
-\Omega_i\Omega_k V_{kj}+\delta_{ij}\delta\Pi\nonumber\\
&-&\pi G\rho\biggl(2B_{ij}V_{ij}
-a_i^2\delta_{ij}\sum_{l=1}^3A_{il}V_{ll}
\biggr),\\
{d^2U_{i;j}\over dt^2}&=&2\epsilon_{ilm}\Omega_m
{dU_{l;j}\over dt}+\Omega^2U_{ij}
-\Omega_i\Omega U_{kj}
+\delta_{ij}\left(\frac{\delta\Pi_S}{f_S}-\frac{\delta\Pi_N}{f_N}\right)
\nonumber\\
&-&2\pi G\rho A_iU_{ij}
-2\Omega\beta_{ik}\frac{d}{dt}U_{k;j}.
\label{eq:AS:relative_virial}
\end{eqnarray}
The first equation is identical to the second order virial equation 
for a normal inviscid fluid. The second equation is specific to 
superfluids and contains all the new modes of relative oscillations 
between the normal fluid and superfluid. 
It is clear that the separation of the oscillation modes in the 
CM and relative modes is the result of the symmetry
of the two-fluid hydrodynamic equations with respect to the 
interchange $\alpha\leftrightarrow\beta$. If this symmetry is 
broken the two classes of modes mix. We shall  consider below 
the effect of the viscosity of normal fluid which by definition 
acts only in the normal component and hence breaks 
the $\alpha\leftrightarrow\beta$ symmetry. 

The second order virial equation for viscous fluids,
quite generally, acquires the term 
\begin{eqnarray} 
{\cal P}_{\alpha, ij} = \int_{V_{\alpha}} P_{\alpha, ij} d{\bf x} , \quad 
P_{\alpha, ik} \equiv \delta_{\alpha , N}
\rho_N \nu \left(\frac{\partial u_{\alpha i}}{\partial x_k} 
+\frac{\partial u_{\alpha, k}}{\partial x_i}-\frac{2}{3}
\frac{\partial u_{\alpha, l}}{\partial x_l}
\delta_{ik}\right),
\end{eqnarray}
which is called the shear-energy tensor; $\nu$ is the kinematic 
viscosity\footnote{We use the same symbol $\nu$ for the kinematic 
viscosity and for the unit vector along the vortex circulation; no 
confusion should arises as the latter quantity does not appear in 
the virial equations.}.
For background states which are stationary and without
internal motions the variation of the stress-energy tensor 
is 
\begin{eqnarray} 
\delta {\cal P}_{\alpha, ij}=\delta_{\alpha, N}
\int_{V_{\alpha}} \rho_N \nu \frac{\partial}{\partial t}
\left(\frac{\partial \xi_{\alpha , i}}{\partial x_k} 
+\frac{\partial \xi_{\alpha , k}}{\partial x_i}
-\frac{2}{3}\frac{\partial \xi_{\alpha l}}{\partial x_l}
\delta_{ik} \right).
\end{eqnarray} 
It is impossible in general to express the variations of  
the stress-energy tensor in terms of the virials $V_{\alpha, i;j}$. 
However, in the low Reynolds-number approximation, this tensor 
can be approximated in a perturbative manner using as the leading 
order approximation the proper solutions for the displacements 
corresponding to the inviscid limit. Since the latter are 
linear functions of the virials, $\xi_{N i} = \sum_{k=1}^3 5V_{N, i;k} x_k
/M_N a_k^2$, with $M_N$ being the mass in the normal fluid, one finds
\begin{eqnarray} 
\delta {\cal P}_{\alpha , ij} = 5\nu\delta_{\alpha , N} \frac{d}{dt}
\left(\frac{V_{\alpha, i;j}}{a_j^2}+\frac{V_{\alpha, j;i}}{a_i^2} \right) 
\end{eqnarray} 
in the incompressible limit. Thus,  the second order virial 
equation which includes the viscosity of the normal matter 
becomes 
\begin{eqnarray}
f_{\alpha}{d^2V_{\alpha,i;j}
\over dt^2}&=&2\epsilon_{ilm}\Omega_mf_{\alpha}
{dV_{\alpha,l;j}\over dt}
+\Omega^2f_{\alpha}V_{\alpha,ij}
-\Omega_i\Omega_kf_{\alpha}V_{\alpha,kj}+\delta_{ij}\delta\Pi_{\alpha}
\nonumber\\
&-&f_{\alpha}\pi G\rho\biggl(2B_{ij}V_{\alpha,ij}
-a_i^2\delta_{ij}\sum_{l=1}^3A_{il}V_{\alpha,ll}
\biggr)
\nonumber\\
&-&a_j^2f_{\alpha} f_{\beta}\pi G\rho\biggl[2A_{ij}(V_{\alpha,ij}
-V_{\beta,ij})+\delta_{ij}
\sum_{l=1}^3A_{il}(V_{\alpha,ll}-V_{\beta,ll})\biggr]
\nonumber\\
&-&{\cal S}_{\alpha\beta}f_{\alpha}\omega_S
\beta_{ik}\frac{d}{dt}
\biggl(V_{\alpha,k;j}-V_{\beta,k;j}\biggr)
-\delta_{\alpha, N} 5\nu f_{\alpha}
\frac{d}{dt}\left(\frac{V_{\alpha,i;j}}{a_j^2}
+\frac{V_{\alpha,j;i}}{a^2_i}\right).\nonumber\\
\label{eq:AS:2nd_order_virial_vis}
\end{eqnarray}
Apart from the last term, the  remaining terms in 
Eq. (\ref{eq:AS:2nd_order_virial_vis}) 
manifestly preserve the symmetry with respect to 
the interchange $\alpha \leftrightarrow \beta$; note that
they might have different parities under this transformation. 
The last term breaks this symmetry as 
the viscosity acts only in the normal fluid. 

\subsection{Third order virial equations}

To obtain the third order virial equation we take
the second moment of Eq. (\ref{eq:AS:euler}) and integrate
over $V_\alpha$:
\begin{eqnarray}
{d\over dt}\biggl(\int_{V_\alpha}{d^3x\rho_\alpha u_{\alpha,i} x_j x_k}
\biggr)&=&2\epsilon_{ilm}\Omega_m\biggl(\int_{V_\alpha}
{d^3x\rho_\alpha u_{\alpha,l}} x_j x_k\biggr)
+\Omega^2I_{\alpha,ijk}
\nonumber\\
&-&\Omega_i\Omega_lI_{\alpha,ljk}+2({\cal T}_{\alpha,ij;k} +{\cal T}_{\alpha,ik;j} )
+\delta_{ij}\Pi_{\alpha,k} +\delta_{ik}\Pi_{\alpha,j}  \nonumber\\
&+&{\cal M}_{\alpha\beta,ijk}
+{\cal F}_{\alpha\beta,ijk},
\label{eq:AS:v3}
\end{eqnarray}
where
\begin{eqnarray}
I_{\alpha,ijk}&\equiv&\int_{V_\alpha}{d^3x\,\rho_\alpha x_ix_jx_k}\\
\Pi_{\alpha,i}&\equiv&\int_{V_\alpha}{d^3x\,p_{\alpha,i}}\\
{\cal T}_{\alpha,ij;k}&\equiv&{1\over 2}\int_{V_\alpha}{d^3x\rho_\alpha
u_{\alpha,i}u_{\alpha,j}x_k}\\
{\cal M}_{\alpha\beta,ijk}&\equiv&
-\int_{V_\alpha}\!\! d^3x  x_jx_k\rho_\alpha
\frac{\partial\phi}{\partial x_i}\\
{\cal F}_{\alpha\beta,ijk}&\equiv&\int_{V_\alpha}{d^3x\,
F_{\alpha\beta,i}}x_jx_k.
\end{eqnarray}
Below, we compute only the perturbations of the tensors in the last
line of Eq. (\ref{eq:AS:v3}), which correspond to the gravitational potential
energy and  the mutual friction; the perturbations of
the remainder terms is computed in analogy to EFE.

For the Eulerian perturbation of the gravitational
potential tensor we have
\begin{eqnarray}
-\delta\int_{V_\alpha}\!\! d^3x  x_jx_k\rho_\alpha
\frac{\partial\phi}{\partial x_i}&=&
-\delta G\int_{V_\alpha} d^3x x_j x_k\rho_\alpha({\bf x})
\Biggl[\int_{V_\alpha}\!\! d^3x^{\prime} \rho_\alpha({\bf x}^{\prime})
\frac{(x_i-x^{\prime}_i)}{\vert{\bf x}-{\bf x}^{\prime}\vert^3}\nonumber\\
&+&
\int_{V_\beta}\!\! d^3x^{\prime}\rho_\beta({\bf x}^{\prime})
\frac{(x_i-x^{\prime}_i)}{\vert{\bf x}-{\bf x}^{\prime}\vert^3}
\Biggr].
\label{eq:AS:gravperturb}
\end{eqnarray}
Assuming $V_\alpha=V$ and $\rho_\alpha=f_\alpha\rho({\bf x})$ in the background
equilibrium, and defining [cf. EFE, Chap. 2, Eqs. (14) and (22)]
\begin{eqnarray}
{\cal B}_{ij}&\equiv& G\int_V d^3x^{\prime}
\rho({\bf x}^{\prime})\frac{(x_i-x^{\prime}_i)(x_j-x^{\prime}_j)}
{\vert{\bf x}-{\bf x}^{\prime}\vert^3},\\
{\cal D}_{ik;j}&\equiv&
G\int_V d^3x^{\prime}\rho({\bf x}^{\prime}) x^{\prime}_j \frac{(x_i-x^{\prime}_i)(x_k-x^{\prime}_k)}
{\vert{\bf x}-{\bf x}^{\prime}\vert^3},
\label{eq:ASbdef}
\end{eqnarray}
one finds for the `self-interaction' term that 
\begin{eqnarray}
&&-\delta\int_{V_\alpha}\!\! d^3x  x_jx_k\rho_\alpha
\frac{\partial\phi_{\alpha}}{\partial x_i}
=-\frac{1}{2}f_{\alpha}^2\int_V d^3x\rho({\bf x}) {\mbox{\boldmath $\xi$}}_{\alpha ,l}
\frac{\partial}{\partial x_l}
\left({\cal B}_{ij} x_k + {\cal D}_{ij;k}\right)\nonumber\\
&-&\frac{1}{2}f_{\alpha}^2\int_V d^3x\rho({\bf x}) {\mbox{\boldmath $\xi$}}_{\alpha ,l}
\frac{\partial}{\partial x_l}
\left({\cal B}_{ik} x_j + {\cal D}_{ik;j}\right)
\equiv f_{\alpha}^2
( \delta {\cal M}_{\alpha ,ij;k}+\delta {\cal M}_{\alpha ,ik;j}) .
\label{eq:ASgravaa2}
\end{eqnarray}
To arrive at the symmetric in the indexes $k,j$ we used the identity
\begin{eqnarray}\label{eq:AS:identity}
{\cal B}_{ij}x_k+{\cal D}_{ik;j}={\cal B}_{ik}x_j+{\cal D}_{ij;k}.
\end{eqnarray}
The perturbation of mutual interaction term 
in Eq. (\ref{eq:AS:gravperturb}), assuming again $V_\alpha=V_\beta=V$
and $\rho_\alpha=f_\alpha\rho({\bf x})$ in the background equilibrium is
\begin{eqnarray}
-\delta\int_{V_\alpha}d^3x x_jx_k\rho_\alpha
\frac{\partial\phi_\beta}{\partial x_i}
&=& -f_{\alpha}f_{\beta}\int_V d^3x\rho({\bf x})
 {\mbox{\boldmath $\xi$}}_{\alpha ,l}({\bf x})
\frac{\partial}{\partial x_l}
\left({\cal B}_{ij} x_k + {\cal D}_{ik;j}\right)\nonumber \\
&-&f_{\alpha}f_{\beta}\int_V d^3x\rho({\bf x})
\left[{\mbox{\boldmath $\xi$}}_{\alpha ,l}({\bf x}) 
- {\mbox{\boldmath $\xi$}}_{\beta ,l}({\bf x})\right]
\frac{\partial^2 {\cal D}_{jk}}{\partial x_l\partial x_i}
\label{eq:ASgravab2}
\end{eqnarray}
where
\begin{eqnarray}
{\cal D}_{jk} = G  \int_{V}\!\! d^3 x^{\prime}
\frac{   \rho({\bf x}^{\prime})x^{\prime}_kx^{\prime}_j}{\vert{\bf x}-{\bf x}^{\prime}\vert} .
\end{eqnarray}
Combining Eqs. (\ref{eq:ASgravaa2})  and  (\ref{eq:ASgravab2}) 
we find
\begin{eqnarray}
-\delta\int_{V_\alpha}\!\! d^3x  x_jx_k\rho_\alpha
\frac{\partial\phi}{\partial x_i}
&=&-f_{\alpha}\int_V d^3x\rho({\bf x}) {\mbox{\boldmath $\xi$}}_{\alpha ,l}
\frac{\partial}{\partial x_l}
\left({\cal B}_{ij} x_k + {\cal D}_{ik;j}\right)
\nonumber \\
&-&f_{\alpha}f_{\beta}\int_V d^3x\rho({\bf x})
\left[{\mbox{\boldmath $\xi$}}_{\alpha ,l}({\bf x})
 - {\mbox{\boldmath $\xi$}}_{\beta ,l}({\bf x})\right]
\frac{\partial^2 {\cal D}_{jk}}{\partial x_l\partial x_i}. 
\end{eqnarray}
For the perturbation of the mutual friction tensor, assuming stationary
background equilibrium, we find
\begin{eqnarray}
 \delta {\cal F}_{\alpha\beta,ijk} & =&
\delta\int_{V_\alpha}{d^3x\, F_{\alpha\beta,i}}x_jx_k. \nonumber\\
&=&-{\cal S}_{\alpha\beta}f_S\int_Vd^3x x_jx_k\rho({\bf x})\omega_S
\beta_{il}\left(\frac{d{\mbox{\boldmath $\xi$}}_{S,l}}{dt}
-\frac{d{\mbox{\boldmath $\xi$}}_{N,l}}{dt}\right).
\end{eqnarray}
Putting together all the terms we arrive at 
the third order virial equation
\begin{eqnarray}
f_{\alpha}{d^2V_{\alpha ,i;jk}\over dt^2}
&=&2\epsilon_{ilm}\Omega_m f_{\alpha}{dV_{\alpha ,l;jk}\over dt}
+\delta_{ij}\delta\Pi_{\alpha ,k}+\delta_{ik}\delta\Pi_{\alpha ,j}\nonumber\\
&+&(\Omega^2\delta_{il}-\Omega_i\Omega_l)f_{\alpha}V_{\alpha ,ljk}
- f_{\alpha}\int_V d^3x\rho({\bf x}) {\mbox{\boldmath $\xi$}}_{\alpha ,l}
\frac{\partial}{\partial x_l}
\left({\cal B}_{ij} x_k + {\cal D}_{ik;j}\right)\nonumber\\
&-&f_{\alpha}f_{\beta}\int_V d^3x\rho({\bf x})
\left[{\mbox{\boldmath $\xi$}}_{\alpha ,l}({\bf x})
 - {\mbox{\boldmath $\xi$}}_{\beta ,l}({\bf x})\right]
\frac{\partial^2}{\partial x_l\partial x_i} {\cal D}_{jk}  \nonumber \\
&-&{\cal S}_{\alpha\beta}f_{S}\omega_S\beta_{il}
\left[{dV_{S ,l;jk}\over dt}-{dV_{N ,l;jk}\over dt}\right],
\label{eq:AS:alphamom2}
\end{eqnarray}
where the symmetric in its indexes third order virial is defined as 
\begin{eqnarray}
V_{\alpha ,ijk}=V_{\alpha ,i;jk}+V_{\alpha ,j;ki}+V_{\alpha ,k;ij}.
\end{eqnarray}
To separate the CM and relative motions of the two fluids
introduce the virials
\begin{eqnarray}
V_{i;jk}\equiv f_SV_{S,i;jk}+f_NV_{N,i;jk}\qquad
U_{i;jk}\equiv V_{S,i;jk}-V_{N,i;jk}.
\end{eqnarray}
The new set of equations in terms of these virials is
\begin{eqnarray}
{d^2V_{i;jk}\over dt^2}
&=&2\epsilon_{ilm}\Omega_m {dV_{l;jk}\over dt}
+\delta_{ij}\delta\Pi_{k}+\delta_{ik}\delta\Pi_{j}
+(\Omega^2\delta_{il}-\Omega_i\Omega_l)V_{ljk}\nonumber\\
&-&\int_V d^3x\rho({\bf x}) \left[f_S{\mbox{\boldmath $\xi$}}_{S,l}
+f_N{\mbox{\boldmath $\xi$}}_{N,l}\right]
\frac{\partial}{\partial x_l}
\left({\cal B}_{ij} x_k + {\cal D}_{ik;j}\right),\nonumber \\
\label{eq:AS:totalmom2}
\end{eqnarray}
and
\begin{eqnarray}
{d^2U_{i;jk}\over dt^2}
&=&\left[2\epsilon_{ilm}\Omega_m
-\left(1+\frac{f_{S}}{f_N}\right)\omega_S\beta_{il}\right]
  {dU_{l;jk}\over dt}
+\delta_{ij}\left(\frac{\delta\Pi_{S ,k}}{f_S}
-\frac{\delta\Pi_{N ,k}}{f_N}\right)\nonumber\\
&+&\delta_{ik}\left(\frac{\delta\Pi_{S ,j}}{f_S}-\frac{\delta\Pi_{N ,j}}{f_N}\right)
+(\Omega^2\delta_{il}-\Omega_i\Omega_l)U_{ljk}- 2\pi G\rho A_iU_{ijk}
\nonumber\\
&-& 2\int_V d^3x\rho({\bf x}) [{\mbox{\boldmath $\xi$}}_{S ,l}
-{\mbox{\boldmath $\xi$}}_{N ,l}]
\frac{\partial}{\partial x_l}
\left({\cal B}_{ij} x_k + {\cal D}_{ik;j} \right).
\label{eq:AS:relmom2bis}
\end{eqnarray}
To obtain the last term in Eq. (\ref{eq:AS:relmom2bis}) 
we used the relations [cf. EFE, Chap. 2, Eqs. (29) and (28)]
\begin{eqnarray}
\frac{\partial {\cal D}_{jk}}{\partial x_i} = {\cal D}_{ji;k}
+x_j\frac{\partial {\cal D}_k}{\partial x_i},\qquad
\frac{\partial{\cal D}_k}{\partial x_i}
= {\cal B}_{ik}+x_k\frac{\partial\phi}{\partial x_i},
\end{eqnarray}
and the explicit expression for the gravitational
potential of an ellipsoid, Eq. (\ref{eq:AS:ellips_phi}). 
The terms in the last 
lines of Eqs. (\ref{eq:AS:totalmom2}) and (\ref{eq:AS:relmom2bis})
can be worked out to a form involving a linear combination of
virials and index symbols, however the present form already makes
clear that they will involve the virials describing the CM 
and relative motions, respectively.

\section{Small amplitude oscillations of superfluid 
          Maclaurin spheroid}

In this section, we specialize our discussion to Maclaurin spheroids,
the  equilibrium figures of a self-gravitating fluid
with two equal semi-major axis, say $a_1$ and $a_2$,
rotating uniformly about the third semi-major axis $a_3$
(i.e. the $x_3$ axis). For these figures in many cases analytical 
results are available. The superfluid oscillations of 
more complicated non-axisymmetric 
figures like the Jacobi and Roche ellipsoids require numerical 
analysis which is beyond the scope of this review (see Ref. \cite{SW2000}).
The sequence of quasi-equilibrium figures of Maclaurin spheroids
can be parameterized by the eccentricity
$\epsilon^2 =1-a_3^2/a_1^2$, with (squared) angular velocity
$\Omega^2 = 2 \epsilon^2 B_{13}$, in units of $(\pi\rho G)^{1/2}$.

Surface deformations related to various modes can be
classified by corresponding terms of the
expansion in surface harmonics labeled by indexes $l,m$.
We shall concentrate below on the 
first and second harmonic surface deformations correspond to $l=1,2$
and  $-1\le m \le 1$, $-2\le m \le 2$ respectively.

\subsection{First order}

If the time-dependence of the Lagrangian displacements is
of the form
\begin{eqnarray}
{\mbox{\boldmath $\xi$}}_{\alpha}(x_i,t) 
= {\mbox{\boldmath $\xi$}}_{\alpha}(x_i)e^{\lambda t},
\label{eq:AS:LAGRANGE}
\end{eqnarray}
then the characteristic equation for the first order relative
oscillation modes becomes
\begin{eqnarray}
\lambda^2 U_i = 2 \epsilon_{ilm}\Omega_m\lambda U_l
+(\Omega^2\delta_{ij}-\Omega_i\Omega_j) U_j - 2 A_i U_i
-2 \Omega \tilde\beta_{ij} \lambda U_j,
\label{eq:AS:monopole2}
\end{eqnarray}
where all frequencies are measured in units $(\pi G \rho)^{1/2}$,
$\tilde\beta_{ij}\equiv (1+f_S/f_N)\beta_{ij}$,  and, since
we assumed no internal motions in the background equilibrium,
$\omega_S =2 \Omega$. The CM oscillations are trivial 
as they can be always eliminated by a transformation to the 
center-of-mass reference frame.

Assume that the ellipsoid is rotating about the $x_3$ axis.
Then, writing Eq. (\ref{eq:AS:monopole2}) in components
we find
\begin{eqnarray}
\label{eq:AS:monoeve1}
\left(\lambda^2 + 2 \Omega\tilde\beta\lambda+ 2 A_1 -\Omega^2\right)U_1
- 2 \Omega(1-\tilde\beta')\lambda U_2 = 0,\\
\label{eq:AS:monoeve2}
\left(\lambda^2 + 2 \Omega\tilde\beta\lambda+ 2 A_2 -\Omega^2\right)U_2
+ 2 \Omega(1-\tilde\beta')\lambda U_1 = 0,\\
\label{eq:AS:monoodd}
\left(\lambda^2+ 2 \Omega\tilde\beta''\lambda + 2 A_3 \right) U_{3} = 0.
\end{eqnarray}
The equations which are even and odd with respect to the index
3 decouple. For the perturbations along $x_3$ the relative 
displacement vanishes, $U_3 =0$.
Eq.  (\ref{eq:AS:monoodd}) (which is odd in index 3) gives,
on writing $\lambda = i \sigma$,
\begin{eqnarray}
\sigma^{\rm odd}_{1,2} = \pm \sqrt{2A_3 - \tilde\beta''^2\Omega^2} + i \tilde\beta''\Omega.
\end{eqnarray}
The first order odd parity oscillations are hence
stable and damped if $\tilde\beta''\le 2A_3/\Omega^2$.
For Maclaurin spheroids $  2A_3/\Omega^2 \ge 5.040$; the lower 
bound corresponds to eccentricity of the ellipsoid $\epsilon = 0.865$.
For Jacobi ellipsoids this minimal 
value is slightly lower,  $2A_3/\Omega^2 =4.148$, and occurs at 
the point of the bifurcation of the Jacobi 
sequence from the Maclaurin sequence where the axis-ratio is 
defined by Cos$^{-1}(a_3/a_1) = 54.48$. Since 
the $\tilde\beta''$-coefficient is the measure of friction along the 
average direction of the vorticity, it is reasonable to assume that 
$\tilde\beta''\ll\tilde\beta, \, \tilde\beta'$; 
and since $\tilde\beta\le 1/2 $ and $\tilde\beta'\le 1$,
we may conclude that the odd modes are stable, unless unphysical 
large values of $\tilde\beta''$ are adopted. 
The characteristic equations for the modes even in index 3 is
\begin{eqnarray}\label{eq:ASFIRST_EVEN}
&&\lambda^4 +   4\Omega\tilde\beta\lambda^3 +
2\left[ (A_1 + A_2) + \Omega^2(1 + 2\tilde\beta^2
               - 4\tilde\beta' + 2\tilde\beta'^2) \right]\lambda^2 
\nonumber\\
&&\hspace{2cm}+ 4\tilde\beta\Omega\,\left(A_1 + A_2 
   -\Omega^2\right)\lambda+ (2A_1-\Omega_2) (2A_2-\Omega_2) =0.
\end{eqnarray}
For Maclaurin spheroids ($A_1=A_2$), upon writing $\lambda = i\sigma$,
the solution becomes
\begin{eqnarray} 
\sigma^{\rm even}_{1,2} =i\tilde\beta\Omega
            \pm\sqrt{2A_1 -\Omega^2[1-\tilde\beta^2-(1-\tilde\beta')^2] }.
\end{eqnarray}
These modes represent stable, damped oscillations since the 
inequality  $1-\tilde\beta^2-(1-\tilde\beta')^2 \le 2A_1/\Omega^2$ is always
fulfilled. Indeed, the left-hand side is always larger than unity, 
while the maximal value of the right-hand side is 1/2. The latter upper 
limit is easy to deduce by minimizing the left-hand side of the inequality 
with respect to $0\le\eta/\rho_S\omega_S\le\infty$ defined 
via the relations (\ref{eq:AS:etas}).

The analysis above shows the principal distinction between the modes
governed by the first and second order virial equations. 
At the first order the dissipation determines the {\it unstable 
configuration} (although it appears that this is never the case for 
realistic values of the friction coefficients). At the 
second order, as we shall see below, the dissipation only determines  
the {\it rate of the instability}, but not the stable configuration
itself.

\subsection{Second order}

Second order harmonic deformations correspond to $l=2$
with five distinct values of $m$, $-2\le m \le 2$.
Again, let us assume time-dependent Lagrangian displacements 
to have the form (\ref{eq:AS:LAGRANGE}).
The characteristic equation for the second order 
oscillation modes become 
\begin{eqnarray}\label{eq:ASSMALL_CM}
\lambda^2 V_{i;j}
&=&2\epsilon_{ilm}\Omega_m \lambda
+\Omega^2V_{ij}
-\Omega_i\Omega_k V_{kj}\nonumber\\
&+&\delta_{ij}\delta\Pi
-\pi G\rho\biggl(2B_{ij}V_{ij}
-a_i^2\delta_{ij}\sum_{l=1}^3A_{il}V_{ll}
\biggr)\nonumber\\
&-& 5\nu f_N\lambda\left(\frac{V_{i;j}}{a_j^2}
+\frac{V_{j;i}}{a^2_i}\right)
+5\nu f_N f_S\lambda\left(\frac{U_{i;j}}{a_j^2}
+\frac{U_{j;i}}{a^2_i}\right)
\label{eq:AS:CM_virial_2},\\
\label{eq:ASSMALL_REL}
\lambda^2 U_{i;j}&=&2\epsilon_{ilm}\Omega_m
\lambda U_{l;j}+\Omega^2U_{ij}-\Omega_i\Omega U_{kj}\nonumber\\
&+&\delta_{ij}\left(\frac{\delta\Pi_S}{f_S}-\frac{\delta\Pi_N}{f_N}\right)
-2\pi G\rho A_iU_{ij}\nonumber\\
& -&2\Omega\tilde\beta_{ik}\lambda U_{k;j}
+ 5\nu \lambda \left(\frac{V_{i;j}}{a_j^2}
+\frac{V_{j;i}}{a^2_i}\right)
-5\nu f_S\lambda \left(\frac{U_{i;j}}{a_j^2}
+\frac{U_{j;i}}{a^2_i}\right),
\label{eq:AS:relative_virial_2}
\end{eqnarray}
where the frequencies are measured in the units $(\pi \rho G)^{1/2}$.
Eqs. (\ref{eq:ASSMALL_CM}) and (\ref{eq:ASSMALL_REL}), which (if written 
in components) constitute a coupled set of 18 equations each, contain
all the second harmonic modes of isolated, incompressible, and 
irrotational superfluid ellipsoids. In the next sections, 
we  concentrate on  solutions of these equations 
for the special case of Maclaurin spheroids, i.e. the case where 
the axial symmetry about the axis of rotation is assumed. 

\subsection{Transverse shear modes ($l=2$, $m =\vert 1\vert$)} 

These modes correspond  to  
surface deformations with $\vert m\vert = 1$
and represent relative shearing of the northern
and southern hemispheres of the ellipsoid. They are 
determined by the eight components 
of the Eqs.  (\ref{eq:AS:CM_virial_2}) and (\ref{eq:AS:relative_virial_2}) 
which are  odd in index 3; i.e.  $V_{3;i}$,  $V_{i;3}$,  $U_{3;i}$
and  $U_{i;3}$, where $i=1,2$. 
The odd equations for the virials describing the CM-motions are
\begin{eqnarray}\label{eq:ASmac_modes:v1.1}
&&\left(\lambda^2 +f_N\nu\lambda + 2 B_{13}\right) V_{13}
-\left(\lambda^2 +\gamma f_N\nu\lambda + 2 B_{13}\right) V_{1;3}
\nonumber\\
&&\hspace{4cm}-f_Nf_S\nu\lambda U_{13} + f_Nf_S\gamma\nu\lambda U_{1;3}  
=0,\\
\label{eq:ASmac_modes:v1.2}
&&\left(\lambda^2 +f_N\nu\lambda + 2 B_{23}\right) V_{23} 
-\left(\lambda^2 +\gamma f_N\nu\lambda + 2 B_{13}\right) V_{2;3}
 \nonumber\\
&&\hspace{4cm}
-f_Nf_S\nu\lambda U_{23} + f_Nf_S\gamma\nu\lambda U_{2;3}  
=0,\\
\label{eq:ASmac_modes:v1.3}
&&
\left(\lambda^2 -\gamma f_N\nu\lambda\right) V_{1;3} 
-2 \Omega\lambda V_{2;3} 
+\left(2 B_{13}-\Omega^2 + f_N\nu\lambda\right) V_{13}
\nonumber\\
&&\hspace{4cm}
+f_Nf_S\gamma\nu\lambda U_{1;3} - f_Nf_S\nu\lambda U_{13}
 =0,\\  
\label{eq:ASmac_modes:v1.4}
&&\left(\lambda^2 -\gamma f_N\nu\lambda\right) V_{2;3} 
+2 \Omega\lambda V_{1;3} 
+\left(2 B_{13}-\Omega^2 + f_N\nu\lambda\right) V_{23} 
\nonumber\\
&&\hspace{4cm}
+f_Nf_S\gamma\nu\lambda U_{2;3} - f_Nf_S\nu\lambda U_{23} =0,
\end{eqnarray}
where  $\gamma\equiv 1 -a_1^2/a_3^2$ and we have redefined
the kinematic viscosity as $\nu' = 5\nu/a_1^2$ and 
dropped the prime in the equations above.  
Relations (\ref{eq:AS:vdef}) were used  
to manipulate the original equations to the form above. 
For the virials describing the relative motions the odd 
parity equations are
\begin{eqnarray}  
\label{eq:ASmac_modes:v1.5}
&&
\left(\lambda^2 +2\Omega\tilde\beta''\lambda 
               +f_S\nu\lambda + 2 A_3\right) U_{13} 
-\left(\lambda^2 +2\Omega\tilde\beta''\lambda 
               +\gamma f_S \nu\lambda\right) U_{1;3}  \nonumber\\
&& \hspace{4cm}     -\nu\lambda V_{13}+\gamma\nu\lambda V_{1;3}=0,\\
\label{eq:ASmac_modes:v1.6}
&&
\left(\lambda^2 +2\Omega\tilde\beta''\lambda 
               +f_S\nu\lambda + 2 A_3\right) U_{23} 
-\left(\lambda^2 +2\Omega\tilde\beta''\lambda 
               +\gamma f_S \nu\lambda\right) U_{2;3}  \nonumber\\
&&\hspace{4cm}
         -\nu\lambda V_{23}+\gamma\nu\lambda V_{2;3}=0,\\
\label{eq:ASmac_modes:v1.7}
&&
\left(\lambda^2 +2\Omega\tilde\beta\lambda 
               -\gamma f_S\nu\lambda\right) U_{1;3}
+(2A_1-\Omega^2 +f_S\nu\lambda) U_{13} 
- 2\Omega (1-\tilde\beta')\lambda U_{2;3} 
\nonumber\\
&&\hspace{4cm}
+\gamma\nu\lambda V_{1;3} -\nu\lambda V_{13} = 0,\\
\label{eq:ASmac_modes:v1.8}
&&\left(\lambda^2 +2\Omega\tilde\beta\lambda 
               -\gamma f_S\nu\lambda\right) U_{2;3}
+(2A_1-\Omega^2 +f_S\nu\lambda) U_{23}
+ 2\Omega (1-\tilde\beta')\lambda U_{1;3} \nonumber\\
&&\hspace{4cm} +\gamma\nu\lambda V_{2;3} -\nu\lambda V_{23} = 0.
\end{eqnarray}
According to the symmetries of the 
original virial equation (\ref{eq:AS:2nd_order_virial_vis}),
the two sets (\ref{eq:ASmac_modes:v1.1})-(\ref{eq:ASmac_modes:v1.4}) 
and  (\ref{eq:ASmac_modes:v1.5})-(\ref{eq:ASmac_modes:v1.8}) decouple in the 
limit $\nu \to 0$, as they should.
The dissipation in the  first set is driven by the viscosity of the normal 
matter; the superfluid contributes to the damping of the CM modes
indirectly, via their coupling 
to the relative oscillation modes. In the second set 
the normal matter viscosity directly renormalizes the mutual friction 
damping time scale ($2\Omega\tilde\beta\to 2\Omega\tilde\beta-\gamma f_S\nu$), 
thus reducing the damping of the relative modes. Note that
this renormalization vanishes for a sphere since then 
$\gamma=0$.
One important feature of each set, which remains preserved when they
are  coupled, is the balance between the tensors describing the 
perturbations of the rotational kinetic energy and gravitational energy;
in the first set only the two-index symbols enter ($B_{ij}$), while in the second one appear only the one-index symbols ($A_i$).
As a results the neutral points (if any)
along a sequence of ellipsoids (parametrized in terms of eccentricity) 
remain unaffected by the coupling 
between the different sets. This implies that as long as there are no 
neutral points for the relative transverse-shear modes in the uncoupled 
case, the conclusion about their stability can not be affected by
the viscosity of the normal component. The CM modes do not show neutral 
points along the Maclaurin sequence and therefore their stability is 
guaranteed.

In the absence of the viscosity the components 
of Eq. (\ref{eq:ASSMALL_CM}), which are odd in
index 3, decouple into two separate sets. The first set for 
virials $V_{ij}$, which describes the CM motions of the 
fluids is identical to the one found in EFE:
\begin{eqnarray}\label{eq:ASmac_modes_Vtr:v1.1}
\left(\lambda^2 + 2 B_{13}\right) V_{13}
-\left(\lambda^2  + 2 B_{13}\right) V_{1;3} &=&0,\\
\label{eq:ASmac_modes_Vtr:v1.2}
\left(\lambda^2 + 2 B_{23}\right) V_{23} 
-\left(\lambda^2 + 2 B_{13}\right) V_{2;3}&=&0,\\
\label{eq:ASmac_modes_Vtr:v1.3}
\lambda^2 V_{1;3} -2 \Omega\lambda V_{2;3} 
+\left(2 B_{13}-\Omega^2\right) V_{13} &=&0,\\  
\label{eq:ASmac_modes_Vtr:v1.4}
\lambda^2 V_{2;3} +2 \Omega\lambda V_{1;3} 
+\left(2 B_{13}-\Omega^2 \right)V_{23} &=&0.
\end{eqnarray}
\begin{figure}[htb] 
\begin{center}
\includegraphics[angle=-90,width=\linewidth]{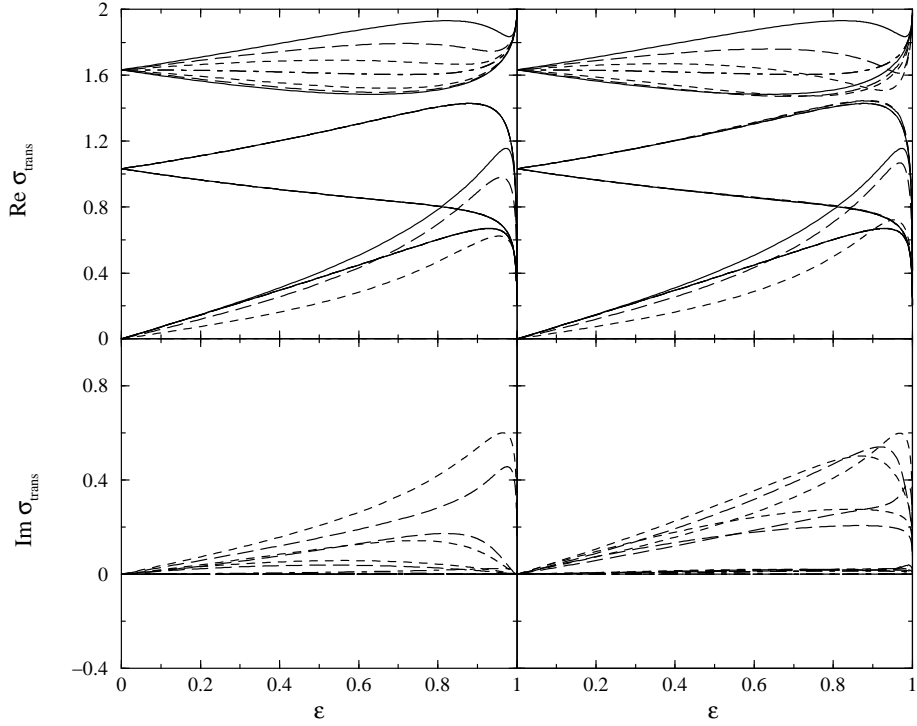}
\end{center}
\caption{
The real (upper panel) and imaginary (lower panel)
parts of the CM and relative transverse-shear modes  of superfluid Maclaurin
spheroid as a function of eccentricity for values of 
$\eta/\omega_S\rho_S =$ 0.0
({\it solid line}), 0.5 ({\it long-dashed line}), 1 ({\it dashed line}), 
and  50  ({\it dashed-dotted line}). 
 The left panel corresponds to the solutions 
in the inviscid limit $\nu =0$. The right panel corresponds to the case
where $\nu = 4 \tilde\beta\Omega $. 
The fraction of the normal fluid $f_N = 0.1$.
}
\label{ASfig:fig1}
\end{figure}
The corresponding modes are described in EFE (see also Fig. 1 below).
The second set, which describes the relative oscillations 
of the fluids,  is 
\begin{eqnarray}  
\label{eq:ASmac_modes_Utr:v1.5}
\left(\lambda^2 +2\Omega\tilde\beta''\lambda + 2 A_3\right) U_{13} 
-\left(\lambda^2 +2\Omega\tilde\beta''\lambda \right) U_{1;3} &=&0,\\
\label{eq:ASmac_modes_Utr:v1.6}
\left(\lambda^2 +2\Omega\tilde\beta''\lambda + 2 A_3\right) U_{23} 
-\left(\lambda^2 +2\Omega\tilde\beta''\lambda \right) U_{2;3} &=&0,\\
\label{eq:ASmac_modes_Utr:v1.7}
\left(\lambda^2 +2\Omega\tilde\beta\lambda \right) U_{1;3}
+(2A_1-\Omega^2) U_{13} - 2\Omega (1-\tilde\beta')\lambda U_{2;3} &=& 0,\\
\label{eq:ASmac_modes_Utr:v1.8}
\left(\lambda^2 +2\Omega\tilde\beta\lambda \right) U_{2;3}
+(2A_1-\Omega^2) U_{23}+ 2\Omega (1-\tilde\beta')\lambda U_{1;3} &=& 0.
\end{eqnarray}
Let us concentrate first on the second set and 
consider the limit of zero mutual friction (i.e. the case where 
the two fluids are coupled only by their mutual gravitational 
attraction).  The characteristic equation can be factorized by 
substituting $\lambda = i\sigma$ to find
\begin{eqnarray}
\sigma\left[\sigma^2 - 2(A_1+A_3) +\Omega^2\right]\pm
2\Omega (\sigma^2 - 2A_3) = 0.
\end{eqnarray}
The purely rotational mode $\sigma=\Omega$
decouples only in the spherical symmetric limit where $A_1 = A_3$.
If only axial symmetry is imposed then the third order
characteristic equation is 
\begin{eqnarray}\label{eq:AS:nondiss_trans}
\sigma^3  \pm 2\Omega{\sigma^2} +
 \left[ -2\,\left(A_1 + A_3\right)  + \Omega^2 \right]\sigma
  \mp 4\,A_3\,\Omega=0.
\end{eqnarray}
It is easy to prove that the modes are always real.
Three complementary modes follow from  Eq. (\ref{eq:AS:nondiss_trans})
via the replacement  $\Omega\to -\Omega$. 
Fig. 1 shows the real and imaginary parts of the
transverse-shear modes along the Maclaurin sequence.  The left panel
shows the results when $\nu = 0$. In that case the relative modes, 
which start for $\Omega\to 0$ at 1.63, are
affected only by the mutual friction (the corresponding 
characteristic equation is of order 6). The CM modes,
which start for $\Omega\to 0$ at 1.03, are unaffected by the 
mutual friction. The modes which 
start for $\Omega\to 0$ at 0, correspond to the rotational frequency of
the spheroid in the low-$\Omega$ limit. The right 
panel shows the same modes but when $\nu / \tilde\beta\Omega =4$.
Interestingly, while the viscous
dissipation considerably affects the relative modes, its effect on the 
real frequencies of the CM modes is marginal.

The damping via mutual friction, as seen from the left panel 
of Fig. 1, is maximal for $\eta/\rho_S\omega = 1$ and decreases to zero 
for $\eta/\rho_S\omega\to 0$ and $\eta/\rho_S\omega\to\infty$. This
behavior is specific to the coupling between the superfluid and the normal
fluid via the vortex state; the communication between these components 
is fastest when the magnitude of the forces on the vortex 
exerted by the superfluid and normal
components are close. In the limiting 
cases the vortices are locked either in the
superfluid ($\eta/\rho_S\omega\to 0$) or the normal 
fluid ($\eta/\rho_S\omega\to \infty$) and hence
the damping is ineffective. 

To conclude, the transverse-shear modes for the relative
and CM modes remain stable along the entire sequence of 
superfluid Maclaurin spheroids.

\subsection{Toroidal modes ($l=2$, $m =\vert 2\vert$)} 

These modes correspond to $\vert m\vert = 2$ and
the motions in this case are confined to planes parallel to the
equatorial plane. 
The toroidal modes are determined by the even in index 3 components
of Eqs.  (\ref{eq:AS:CM_virial_2}) and (\ref{eq:AS:relative_virial_2})
for the virials  $V_{i;i}$,  $V_{i;j}$,  $U_{i;i}$ and  $U_{i;j}$, 
where $i,j=1,2$. 
\begin{figure}[htb] 
\begin{center}
\includegraphics[angle=-90,width=\linewidth]{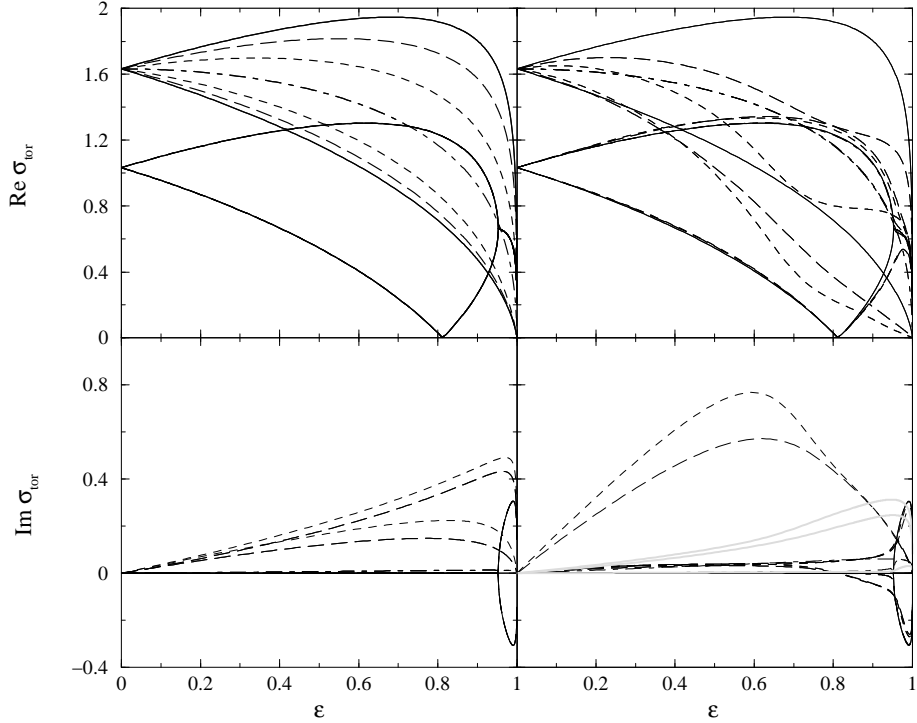}
\end{center}
\caption{
The CM and relative toroidal modes of  
superfluid Maclaurin spheroid. 
The imaginary parts shown in {\it grey}
are magnified by factor 10. Other conventions  are the same as in Fig. 1.
Note the secular instability at the bifurcation point $\epsilon = 0.81$ 
and the dynamical instability at the point $\epsilon = 0.93 $.
}
\label{ASfig:fig2}
\end{figure}
These equations can be manipulated to a 
set of four equations, which read
\begin{eqnarray}
&&\left(\lambda^2 + 2f_N\nu\lambda + 4B_{12} -2\Omega^2\right) V_{12}
+\Omega\lambda (V_{11}-V_{22})\nonumber\\
&&\hspace{4cm} - 2f_Nf_S\nu\lambda U_{12} =0,\\
&&\left(\lambda^2 + 2f_N\nu\lambda + 4B_{12} -2\Omega^2\right) (V_{11}-V_{22})
\nonumber\\
&&\hspace{4cm}-4\Omega\lambda V_{12}- 2f_Nf_S\nu\lambda (U_{11}-U_{22}) =0,\\
&&\left(\lambda^2+2\Omega\tilde\beta\lambda+2f_S\nu\lambda
+4A_1-2\Omega^2\right)U_{12}\nonumber\\
&&\hspace{4cm}+\Omega(1-\tilde\beta')\lambda(U_{11}-U_{22})-2\nu\lambda V_{12}=0,\\
&&\left(\lambda^2+2\Omega\tilde\beta\lambda+2f_S\nu\lambda+4A_1-2\Omega^2\right)
(U_{11}-U_{22})\nonumber\\
&&\hspace{4cm}-4\Omega(1-\tilde\beta')\lambda U_{12}
-2\nu\lambda (V_{11}-V_{22})=0.
\end{eqnarray}
In the inviscid limit equations above decouple into separate 
sets for the CM and relative oscillations. The CM oscillations 
are described by the equations
\begin{eqnarray}
&&\left(\lambda^2 + 4B_{12} -2\Omega^2\right) V_{12}
+\Omega\lambda (V_{11}-V_{22})= 0, \\
&&\left(\lambda^2+ 4B_{12} -2\Omega^2\right) (V_{11}-V_{22})
-4\Omega\lambda V_{12}=0,
\end{eqnarray}
and their solutions are documented in EFE. The relative 
oscillations are described by the following equations
\begin{eqnarray}  \label{eq:ASmac_modes:2.6}
\left(\lambda^2 + 2 \Omega \tilde\beta \lambda+4 A_1- 2 \Omega^2 \right)(U_{11}-U_{22})
-4 \Omega \lambda (1-\tilde\beta') U_{12} &=& 0, \\\
     \label{eq:ASmac_modes:2.7}
 \left(\lambda^2 + 2 \Omega \tilde\beta \lambda+4 A_1- 2 \Omega^2\right) U_{12}
+\Omega\lambda (1-\tilde\beta')(U_{11}-U_{22}) &=& 0 .
\end{eqnarray}
and the characteristic equation for the relative toroidal modes is:
\begin{eqnarray}  \label{eq:ASmac_modes:2.8}
&&\lambda^4 + 4 \tilde\beta\Omega \lambda^3+
    (8 A_1 + 4 \tilde\beta^2\Omega^2 
- 8\tilde\beta'\Omega^2 + 4\tilde\beta'^2\Omega^2)\lambda^2
   \nonumber \\
   &&\hspace{2cm}+
   8\tilde\beta\Omega (2A_2-\Omega^2) \lambda
   +4 (2A_1-\Omega^2)^2 = 0.
\end{eqnarray}
In the frictionless limit the modes can be found analytically from
\begin{eqnarray}
(\lambda^2 + 4 A_1 - 2 \Omega^2)^2 + 4 \Omega^2\lambda^2  =0,
\end{eqnarray}
which is factorized by writing $\lambda=i\sigma$.
The two solutions are then
\begin{eqnarray}\label{eq:AS:sig_tor}
\sigma_{1,2}=\Omega\pm\sqrt{4 A_1 - \Omega^2}.
\end{eqnarray}
and there are
two complementary modes which are found by 
substituting $-\Omega$ for $\Omega$.
If the mutual friction is included the characteristic equation
describing  the relative oscillation modes 
is of order 4; in the presence of viscosity of the normal component, 
again the CM and relative oscillation modes couple and
the characteristic equation is of order 8.
The real and imaginary parts of the dissipative toroidal modes are shown
in Fig. 2, for the same values of parameters as in Fig. 1. 
As for the transverse-shear modes,
the CM and relative modes start at 1.03 and 
1.63, respectively, when $\Omega\to 0$. 
In the inviscid limit (left panel), the CM modes are unaffected, 
as they should, while the relative modes are driven against each 
other and merge in the limit of strong coupling.
The damping of the relative modes is finite, while for the CM modes
it vanishes exactly up to the point of the onset of the dynamical 
instability at $\epsilon =0.95$; beyond this point a mode becomes 
dynamically (i.e. in the absence of dissipation) unstable.
If the kinematic viscosity is finite (right panel), the real 
parts of the relative modes are strongly affected, while the effect
of the viscosity on the CM modes is marginal. 
The imaginary part 
of a CM mode changes its sign at the  bifurcation point, where
$2 B_{12} =\Omega^2$ and $\epsilon = 0.813$. This signals the onset 
of the classical secular instability of the Maclaurin spheroid.
The new feature here is that the mutual friction contributes 
to the secular instability of a CM-mode. On the other hand, ordinary
viscosity does not drive the relative modes unstable, in agreement with the 
fact that there are no neutral points for these modes along the 
entire Maclaurin sequence. One may conclude that the agents which break 
the superfluid/normal fluid symmetry can not cause an instability 
of the relative modes. The only possibility that the relative modes
become unstable is a shift of the balance
between the kinetic and potential energy perturbations, which might
occur for compressible fluids, e.g. polytrops. This problem will be
studied elsewhere.

\subsection{Pulsation mode ($l=2$, $m=0$)}

Pulsation modes (or breathing modes) are the generalization of the
radial pulsation modes of a sphere to the case of rotation. 
They  correspond to $l = 2$ and $m=0$ indexes in the expansion
in spherical harmonics. 
The pulsation modes are determined by the full set of equations
which are even in index 3. By suitable combination 
of the equations for the virials  $V_{i;i}$,  $V_{i;j}$,  $U_{i;i}$ 
and  $U_{i;j}$, where $i=1,2,3$ and $j=1,2$ the original set of  
equations can be reduced to 
\begin{eqnarray} 
\left(\lambda^2/2+f_N\nu\lambda+4B_{11}-2B_{13}-\Omega^2\right)
             \left(V_{11}+V_{22}\right)\nonumber\\
-\Bigl[\lambda^2 +2f_N\nu (1-\gamma)\lambda
+ 6B_{33} - 2 B_{13}\Bigr] V_{33}
\nonumber\\
+2\Omega\lambda(V_{1;2}-V_{2;1})
-f_Nf_S\nu\lambda(U_{11}+U_{22})
+2f_Nf_S\nu (1-\gamma) U_{33} 
=0,\\
\lambda^2 (V_{1;2}-V_{2;1}) -\Omega\lambda (V_{11}+V_{22})
=0,\\
\left(\lambda^2/2+2A_1-\Omega^2 
+f_S\nu\lambda +\Omega\tilde\beta \lambda\right)
(U_{11}+U_{22})\nonumber\\ 
-\Bigl[\lambda^2 + 2 \Omega\tilde\beta'' \lambda
+ 4 A_3 + 2f_S \nu (1-\gamma)\lambda
\Bigr] U_{33}\nonumber\\ 
+ 2\Omega(1-\tilde\beta')\lambda (U_{1;2}-U_{2;1})
 -\nu \lambda (V_{11}+V_{22})  +2\nu (1-\gamma)\lambda V_{33}
=0,\\
(\lambda^2 + 2 \Omega \tilde\beta \lambda) (U_{1;2}-U_{2;1}) 
-\Omega(1-\tilde\beta')\lambda (U_{11}+U_{22})
=0.
\end{eqnarray} 
The characteristic equation is found by supplementing these equations
by the  divergence free constraint on the virials of the CM and relative 
motions:
\begin{eqnarray} \label{eq:ASDIV}
\sum_{i=1}^3\frac{V_{ii}}{a_i^2} = 0,\quad
\sum_{i=1}^3\frac{U_{ii}}{a_i^2} = 0.
\end{eqnarray}
An equivalent form of the divergence free
constraint for Maclaurin spheroids can be written in terms of eccentricity,
$(V_{11}+V_{22})(1-\epsilon^2)  + V_{33} = 0$ and similarly for 
$U_{ii}$.

In  absence of viscosity the equations above decouple into 
two independent sets for CM and relative oscillations. The 
CM oscillations are described by 
\begin{eqnarray}
&&\left(\lambda^2/2+4B_{11}-2B_{13}-\Omega^2\right)
             \left(V_{11}+V_{22}\right)\nonumber\\
&&\hspace{2cm}-\Bigl(\lambda^2+ 6B_{33} - 2 B_{13}\Bigr) V_{33}
+2\Omega\lambda(V_{1;2}-V_{2;1}) = 0,\\
&&\lambda^2 (V_{1;2}-V_{2;1}) -\Omega\lambda (V_{11}+V_{22}) =0,
\end{eqnarray}
and coincide with the pulsation modes treated in EFE. The relative 
oscillation modes are defined by the equations
\begin{eqnarray} \label{eq:ASmac_modes:3.1}
&&\left(\lambda^2/2+ \Omega\tilde\beta\lambda-\Omega^2+2 A_1\right)(U_{11}+U_{22})  
     \nonumber \\
&&\hspace{1cm}+ 2 \Omega\lambda (1-\tilde\beta')(U_{1;2}-U_{2;1})
-(\lambda^2 + 4 A_3+2\Omega \tilde\beta'' \lambda) U_{33}=0,\\
    \label{eq:ASmac_modes:3.2}
&& \left(\lambda^2 + 2 \Omega\tilde\beta\lambda\right) (U_{1;2}-U_{2;1})
-\Omega\lambda(1-\tilde\beta') (U_{11}+U_{22}) = 0,
\end{eqnarray}
which can be combined to:
\begin{eqnarray}\label{eq:ASmac_modes:3.3}
&&\left[\left(\lambda^2+2\Omega\tilde\beta\lambda - 2 \Omega^2 + 4 A_1\right)
(\lambda^2+2 \Omega\tilde\beta \lambda) + 4\Omega^2\lambda^2
(1-\tilde\beta')^2\right](U_{11}+U_{22})\nonumber \\
&&\hspace{2cm}-2\left[\left(\lambda^2+2\Omega\tilde\beta\lambda\right)
\left(\lambda^2+2\Omega\lambda\tilde\beta''+4A_3\right)\right]U_{33}=0.
\end{eqnarray}
The modes are found by supplementing this equation
by the  divergence free condition  (\ref{eq:ASDIV}).
The third order characteristic equation in the inviscid limit is 
\begin{eqnarray}
&&(3 - 2\epsilon^2)\lambda^3 +
    [8\tilde\beta\Omega + 4\tilde\beta''\Omega
    - 4(\tilde\beta+\tilde\beta'')\epsilon^2\Omega]\lambda^2+
   [4A_1 + 8(1 - \epsilon^2)A_3\nonumber \\
&&\hspace{2cm} + 2\Omega^2
   + 4\tilde\beta^2\Omega^2 - 8\tilde\beta'\Omega^2 +
      4\tilde\beta'^2\Omega^2 
+ 8(1-\epsilon^2)\tilde\beta\tilde\beta''\Omega^2 ] \lambda \nonumber \\
&&\hspace{2cm}+8A_1\tilde\beta\Omega +
16(1 - \epsilon^2)A_3\tilde\beta\Omega 
- 4\tilde\beta\Omega^3 = 0,
\end{eqnarray}
where the trivial mode $\lambda = 0$ is neglected.
In the frictionless limit we find ($\lambda=i\sigma$ as before)
\begin{eqnarray}\label{eq:AS:sig_pul_nondiss}
\sigma = \pm\left[
\frac{ 2 \Omega^2 + 4 A_1 + 8 A_3 (1-\epsilon^2)}{(3 - 2 \epsilon^2) }
\right]^{1/2} .
\end{eqnarray}
\begin{figure}[htb] 
\begin{center}
\includegraphics[angle=-90,width=\linewidth]{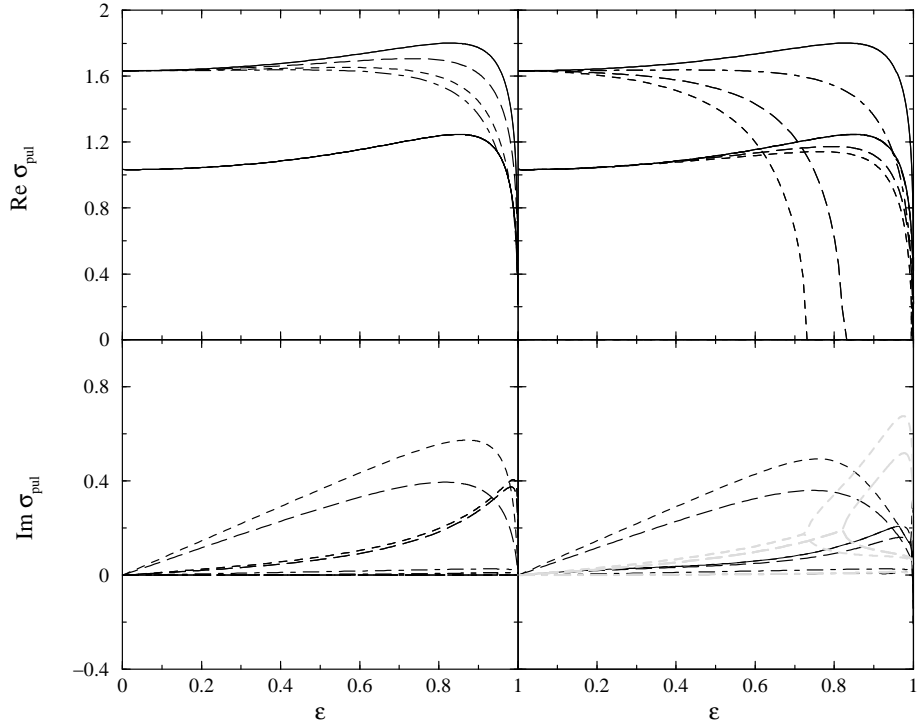}
\end{center}
\caption{
The CM and relative pulsation modes of  
superfluid Maclaurin spheroid.  Conventions  are the same as in Fig. 1
and 2.
}
\label{ASfig:fig3}
\end{figure}
The pulsation modes for a sphere follow in the limit
($\epsilon, \Omega) \to 0$:
for a sphere  $A_i/(\pi\rho G) = 2/3$, and Eq. (\ref{eq:AS:sig_pul_nondiss})
reduces to $\sigma^2 = 8/3$  [$\sigma$ is given
in units of $(\pi\rho G)^{1/2}$].
This result could be compared with the pulsation modes of an
ordinary sphere:  $\sigma^2 = 16/15$. Thus a superfluid sphere,
apart form the ordinary pulsations, shows pulsations
at frequencies roughly twice as large as the ordinary ones.
In the general case where the  viscosity of the normal fluid is 
taken into account 
the characteristic equation is of fifth order. The real and 
imaginary parts of the roots are shown in Fig. 3.
In the inviscid limit the CM modes are again unaffected, while the 
relative modes are suppressed by the mutual friction. The damping 
of these modes is maximal when $\eta/\rho_S\omega =1$ 
 and the motions correspond to stable, damped oscillations.
In the presence of viscosity, the relative modes are strongly damped 
and eventually become neutral. The CM modes are weakly affected. The 
imaginary parts remain always positive, i.e. one finds stable, damped 
oscillations. Note that at the point where a relative pulsation 
mode becomes neutral 
the number of the imaginary components increases by one (as generally
expected for the roots of polynomials), but there are no changes in the 
sign of the imaginary part of the neutral mode.

\section{Summary}

Let us briefly summarize the main qualitative features 
of the oscillations of superfluid self-gravitating systems \cite{SW2000}:
\begin{itemize}
\item The oscillation modes of the 
      superfluid ellipsoids separate into
      two generic classes which correspond to  {\it co-moving} 
      and {\it relative oscillations}. The
      oscillation frequencies of these two classes have distinct 
      values in the both slow and rapid rotation limits.
      The  first class  of oscillations is identical to those
      of classical single-fluid ellipsoids in the incompressible 
      and inviscid limits.
      Corresponding modes are undamped if the Euler equations
      of fluids  are symmetric/anti-symmetric with respect to the
      interchange  $\alpha\leftrightarrow \beta$. When the fluids 
      are coupled  by mutual friction  
      and mutual gravitational attraction this symmetry is 
      preserved. 

\item  The second class of oscillations, which is new,  corresponds
       to relative 
       motions of the fluids. The modes
       are damped by the mutual friction between the superfluid and 
       the normal fluid. These modes correspond to stable 
       oscillations.

\item The co-moving (CM) modes emit gravitational 
       radiation and undergo radiation 
      reaction instabilities in full analogy to single-fluid 
      ellipsoids \cite{GRAVRAD}. 
      The relative modes do not emit gravitational radiation 
      at all, since the mass current associated with them is zero.
      This picture must hold true for a  more general class of
      Chandrasekhar-Friedman-Schutz (CFS) radiation reaction 
      instabilities, which are intrinsic to self-gravitating 
      Newtonian fluids  \cite{FS}.
           
\item If the $\alpha\leftrightarrow \beta$ symmetry is broken 
      the two classes of modes mix, for example, 
      when the normal fluid is viscous. 
      The main effect of the mixing is the {\it renormalization of the
      mutual friction and viscosity}. The relative modes remain stable 
      as there are no distinct neutral points for these modes along the 
      ellipsoidal sequences. The CM modes become unstable
      at the classical points of onset of secular/dynamical instabilities, 
      for example at the point of bifurcation of the 
      Maclaurin spheroid into a Jacobi ellipsoid.
\end{itemize}
These  qualitative features  are based on general symmetries of 
underlying hydrodynamic equations {\it and} 
the conditions of equilibrium of self-gravitating fluids 
which are independent of the superfluid nature of underlying 
fluids (e.g. the existence of bifurcation  points). Therefore we 
may conclude that these features will be preserved in more complex models
of oscillations of superfluid neutron stars.

\section*{Acknowledgments}

This work was supported in part by FOM 
at KVI (Groningen) and by CNRS at IPN (Orsay), and by NASA at Cornell.

\end{document}